\documentclass{article}

\PassOptionsToPackage{numbers, compress}{natbib}
\usepackage[final]{neurips_2024}

\usepackage[utf8]{inputenc} 
\usepackage[T1]{fontenc}    
\usepackage{microtype}
\usepackage{graphicx}
\usepackage{subcaption}

\usepackage{booktabs} 

\usepackage{hyperref}

\usepackage{amsmath}
\usepackage{amssymb}

\usepackage{multirow}
\usepackage{multicol}
\usepackage{mathtools}
\usepackage{amsthm}

\usepackage{xspace}
\usepackage[capitalize,noabbrev]{cleveref}
\usepackage{colortbl}
\theoremstyle{plain}

\usepackage{subcaption}
\theoremstyle{definition}

\theoremstyle{remark}

\usepackage[textsize=tiny]{todonotes}

\usepackage{tcolorbox} 
\usepackage{listings} 

\usepackage[frozencache,cachedir=minted-cache]{minted}
\usepackage{xcolor} 
\usepackage{caption}
\usepackage{enumitem}

\definecolor{lightorange}{RGB}{253,188,180}
\definecolor{lightblue}{RGB}{180,208,253}
\definecolor{problemcolor}{RGB}{153,0,153}
\definecolor{modelcolor}{RGB}{0, 0, 255}

\definecolor{pythonblue}{RGB}{84, 184, 255}
\definecolor{soappink}{RGB}{255, 163, 203}

\newtcolorbox{problembox}[1][]{
    colback=soappink!5!white,
    colframe=soappink!75!black,
    title=\textbf{Task Description},
    top=2mm,
    bottom=2mm,
    left=2mm,
    right=2mm,
    arc=2mm,
    boxsep=1mm,
    boxrule=0mm,
    #1
}

\newtcolorbox{modelbox}[1][]{
    colback=pythonblue!5!white,
    colframe=pythonblue!75!black,
    title=\textbf{Model},
    top=2mm,
    bottom=2mm,
    left=2mm,
    right=2mm,
    arc=2mm,
    boxsep=1mm,
    boxrule=0mm,
    #1
}

\setminted{breaklines}    
\setminted{fontsize=\scriptsize} 

\newcommand{\name}{\textsc{Effi-Learner}\xspace}

\title{\name: Enhancing Efficiency of \\  Generated Code via Self-Optimization}
\author{%
  Dong Huang\thanks{Equal Contribution.} \\
  The University of Hong Kong\\
  \texttt{dhuang@cs.hku.hk} \\
  \And
  Jianbo Dai\footnotemark[1]\\
  University of Edinburgh \\
  \texttt{j6dj6d@gmail.com} \\
  \And
  Han Weng \\
  Beijing University of \\
  Posts and Telecommunications \\
  \texttt{han.weng@bupt.edu.cn} \\
  \And
  Puzhen Wu \\
  University College Dublin \\
  \texttt{puzhen.wu@ucdconnect.ie} \\
  \And
  Yuhao Qing \\
  The University of Hong Kong \\
  \texttt{yhqing@cs.hku.hk} \\
  \And
  Heming Cui \\
  The University of Hong Kong \\
  Shanghai AI Laboratory\\
  \texttt{heming@cs.hku.hk} \\
  \And
  Zhijiang Guo\thanks{Corresponding Author.}  \\
  University of Cambridge \\
  \texttt{zg283@cam.ac.uk} \\
  \And
  Jie M. Zhang \\
  King’s College London \\
  \texttt{jie.zhang@kcl.ac.uk} \\
}

\begin{document}

\maketitle


\begin{abstract}

Large language models (LLMs) have shown remarkable progress in code generation, but their generated code often suffers from inefficiency, resulting in longer execution times and higher memory consumption. To address this issue, we propose \textbf{\name}, a self-optimization framework that utilizes execution overhead profiles to improve the efficiency of LLM-generated code. \name first generates code using an LLM, then executes it locally to capture execution time and memory usage profiles. These profiles are fed back to the LLM, which then revises the code to reduce overhead. To evaluate the effectiveness of \name, we conduct extensive experiments on the EffiBench, HumanEval, and MBPP with 16 open-source and 6 closed-source models. Our evaluation results demonstrate that through iterative self-optimization, \name significantly enhances the efficiency of LLM-generated code. For example, the execution time (ET) of StarCoder2-15B for the EffiBench decreases from 0.93 (s) to 0.12 (s) which reduces 87.1\%  execution time requirement compared with the initial code. The total memory usage (TMU) of StarCoder2-15B also decreases from 22.02 (Mb*s) to 2.03 (Mb*s), which decreases 90.8\% total memory consumption during the execution process. The source code of \name was released in \url{https://github.com/huangd1999/EffiLearner}.

\end{abstract}
\section{Introduction}
\label{intro}

\begin{figure*}[htbp]
    \centering
    \begin{minipage}{1\textwidth}
        \begin{problembox}
            \raggedright
            \textbf{Problem:} Given a non-empty array of integers nums, every element appears twice except for one. Find that single one. You must implement a solution with a linear runtime complexity and use only constant extra space.
            \noindent
            \begin{minipage}{0.5\textwidth}
                \textbf{Example:}
                \textbf{Input:} nums = [4,1,2,1,2]
            \end{minipage}%
            \begin{minipage}{0.5\textwidth}
                \hfill \textbf{Output:} 4
            \end{minipage}
\begin{minted}{python}
assert singleNumber([4, 1, 2, 1, 2]) == 4
\end{minted}
        \end{problembox}
    \end{minipage}
        \hfill
        \begin{minipage}[t]{0.5\textwidth}
            \begin{modelbox}[title=\textbf{Initial Completion with Profile}]
\begin{minted}{python}
Timer unit: 1e-06 s
Total time: 0.0312331 s
Mem unit: MiB
Time   Mem usage    Line Contents 
================================================
            def singleNumber(nums):
  371.2 50.9    for i in range(len(nums)):
30860.9 50.9        if nums.count(nums[i]) == 1:
    0.9 50.9            return nums[i]
\end{minted}
            \end{modelbox}
        \end{minipage}%
        \begin{minipage}[t]{0.5\textwidth}
            \begin{modelbox}[title=\textbf{\name Completion with Profile}]
\begin{minted}{python}
Timer unit: 1e-06 s
Total time: 0.000112618 s
Mem unit: MiB
Time  Mem usage   Line Contents
==============================================
          def singleNumber(nums):
112.6 50.7    return reduce(operator.xor, nums)


\end{minted}
            \end{modelbox}
        \end{minipage}%
        \hfill
    \caption{A case for the task with code and \name refined version. The lower left panel shows the initial completion generated by an LLM, its profile shows its inefficiency. The lower right panel shows the final efficient answer output by applying \name.}
    \vspace{-0.5cm}
    \label{fig:example}
\end{figure*}

Large language models (LLMs) have recently achieved significant advancements across various tasks~\citep{GPT35turbo,Gemini23,Claude3,Llama3}. LLMs such as GPT-4~\citep{GPT4} and Copilot~\citep{Copilot} have demonstrated considerable efficacy in code-related applications, including code completion~\citep{ChenCodex2021,Austin2021}, debuggingg~\citep{Haque2022,ChenSelfDebug23}, and translation~\citep{RoziereLCL20,AhmadTCC23}. These innovative tools have been seamlessly integrated into popular development environments, significantly augmenting developer productivity by providing intelligent code recommendations based on natural language instructions. However, the primary focus of existing efforts has predominantly been on the correctness of the generated code \cite{fan2023large}, ensuring it meets the functional requirements and adheres to syntactical norms. 

Despite advancements in ensuring code correctness, there remains a significant gap in the literature regarding the efficiency of code produced by LLMs. Efficiency is crucial as it translates to faster execution and lower memory and processing power consumption, which is especially important in resource-constrained environments such as mobile devices or embedded systems \cite{fan2023large,qiu2024efficient,coignion2024performance,niu2024evaluating,du2024mercury,shi2024efficient,capra2012software,mancebo2021process,pereira2021ranking}. Recent studies~\citep{shi2024efficient, niu2024evaluating,liu2024evaluating,huang2024effibench,pie_iclr_2024_spotlight,waghjale2024ecco} reveal that LLM-generated code often exhibits lower efficiency in terms of execution time and memory usage when compared to canonical solutions. For instance, on a benchmark that evaluates efficiency, EffiBench~\citep{huang2024effibench}, even the most powerful LLM (e.g., GPT-4-Turbo) generates code with suboptimal efficiency. The average and worst-case execution times are 1.69 and 45.49 times longer than those of the canonical solutions, respectively. This inefficiency underscores the need for developing new methods focused on evaluating and improving the efficiency of code generated by LLMs, ensuring that they produce correct and highly efficient code.

To bridge this gap, we draw inspiration from the methodology used by coders on coding platforms. When addressing a programming problem, coders typically write an initial program that is executable on the test cases. Next, they execute the code and obtain a profile of the execution time and memory usage overhead, as shown in Figure~\ref{fig:example}. Based on this overhead profile, the coder optimizes the code to enhance its efficiency. During this process, the coder extracts key information (e.g., execution times and memory usage of each line) from the overhead profile, which helps identify lines or operators that require significant overhead (e.g., loops that execute multiple times or unnecessary variables being saved). This information assists the coder in optimizing their code.

With this motivation, we propose \textbf{\name} to improve the efficiency of LLM-generated code. As shown in Figure~\ref{fig:framework},  \name first requires the LLM to generate code based on the task description. Then, \name executes the generated code locally and captures the execution time and memory usage profile. These overhead profiles are fed back into the LLM, which then revises the code to reduce the overhead. Through multi-iteration self-optimization, the efficiency of the LLM-generated code is improved.  While it's true that the iterative process requires extra time, it's crucial to recognize the long-term advantages that come with this investment. By optimizing the code, we can enhance the overall efficiency once it's deployed.

To evaluate the effectiveness of \name, we conduct extensive experiments on the EffiBench and two commonly used code generation benchmarks (i.e. HumanEval~\citep{ChenCodex2021} and MBPP~\citep{Austin2021}) with 16 open-source and 6 closed-source models. We compare the efficiency of the code generated by the LLM before and after applying \name. The experimental results demonstrate that \name significantly improves the efficiency of the LLM-generated code. For example, the execution time (ET) of StarCoder2-15B decreases from 0.93 (s) to 0.12 (s) which reduces 87.1\%  execution time requirement compared with the initial code. The max memory usage (MU) of DeepSeek-6.7B-Ins also decreased from 259.73 (Mb) to 36.97 (Mb), which reduces 85.8\% max memory consumption for the code execution requirement. The total memory usage (TMU) of StarCoder2-15B also decreases from 22.02 (Mb*s) to 2.03 (Mb*s), which decreases 90.8\% total memory consumption during the execution process.

\section{Related Work}
\label{related}

\subsection{LLMs for Code Generation}
The increasing popularity of LLMs for code generation has coincided with the growing availability of open-source code repositories and the need to boost developer productivity. Initial efforts focused on training models specifically for coding tasks, such as CodeT5~\citep{WangCodeT52021}, AlphaCode~\citep{Lialphacode2022}, CodeGen~\citep{NijkampPHTWZSX23}, InCoder~\citep{FriedAL0WSZYZL23}, StarCoder~\citep{LiStarCoder203}, SantaCoder~\citep{Loubnasanta2023} and DeepSeek Coder~\citep{deepseekcoder}. Contrastingly, models such as Codex~\citep{ChenCodex2021} and CodeLLaMA~\citep{Roziere2023} represent a subsequent stride, being finetuned from foundation models~\citep{BrownMRSKDNSSAA20,Touvron2023}.  These code LLMs have been applied to various tasks, including code generation~\citep{ChenCodex2021, Dai2024mhpp}, program repair~\citep{Haque2022,JiangLLT23}, automated testing~\citep{LemieuxILS23,Deng2023}, code translation~\citep{RoziereLCL20,AhmadTCC23}, type prediction~\citep{MirLPG22,WeiDD23}, and code summarization~\citep{HasanMIMHHAIS21,AhmedD22}. Among these, code generation, where models generate code snippets based on natural language descriptions or docstrings, has become a critical domain for evaluating LLMs. While LLMs have achieved impressive results in code generation tasks like HumanEval~\citep{ChenCodex2021} and MBPP~\citep{Austin2021}, their efficiency has received less attention. Recent studies~\citep{shi2024efficient, huang2024effibench, niu2024evaluating} have shown that LLM-generated code exhibits lower efficiency in execution time and memory usage compared to canonical solutions. 
These findings highlight the need for further research and development to improve the efficiency of LLM-generated code. In this work, we propose the first method that significantly improves the efficiency of code generated by a wide range of LLMs.

\subsection{Learning From Feedback}
A prevalent strategy for improving the behavior of LLMs is learning from feedback, mirroring human learning where individuals refine their actions through trial, error, and correction~\citep{boyd1983reflective,metcalfe2017learning}. Early efforts involve using human feedback to evaluate and refine models~\citep{KreutzerKMR18,Ouyang0JAWMZASR22,Glaese2022}. To minimize human intervention, another strategy focuses on automated feedback. These methods iteratively learn from automatically generated feedback signals, understanding the consequences of their actions and adapting their behaviors. The source of this automated feedback can be diverse, ranging from the LLM itself~\citep{MadaanSelfRefine23, ShinnReflexion23}, external tools~\citep{GouCritic2023,Lu2024pda} or verifiers~\citep{Lu2024Auto}, external knowledge sources~\citep{GaoRARR23, YuReFeed23} and even generation logits~\citep{Yao2024learning}. In code generation, the program executor is frequently used as a source of feedback for refining the model's initial code. For example, Self-Edit~\citep{ZhangSelfEdit23} and Self-Evolve~\citep{JiangSelfEvolve23} execute the initial program on example test cases and provide the execution results as feedback, prompting the LLM to refine the code. Self-Debug~\citep{ChenSelfDebug23} explores using program explanation, unit tests, and program interpreters as feedback types. ALGO~\citep{ZhangAlgo23} employs a more fine-grained approach by generating a reference oracle program that solves the problem with an exhaustive search. Feedback is then collected by comparing the generated outputs with the oracle. While existing work primarily focuses on using feedback to edit the initial code to ensure correctness, our method explores using overhead profiles to improve the efficiency of the code.

\section{\name}
\label{framework}

\begin{figure*}[t]
    \centering
    \includegraphics[width=0.8\textwidth]{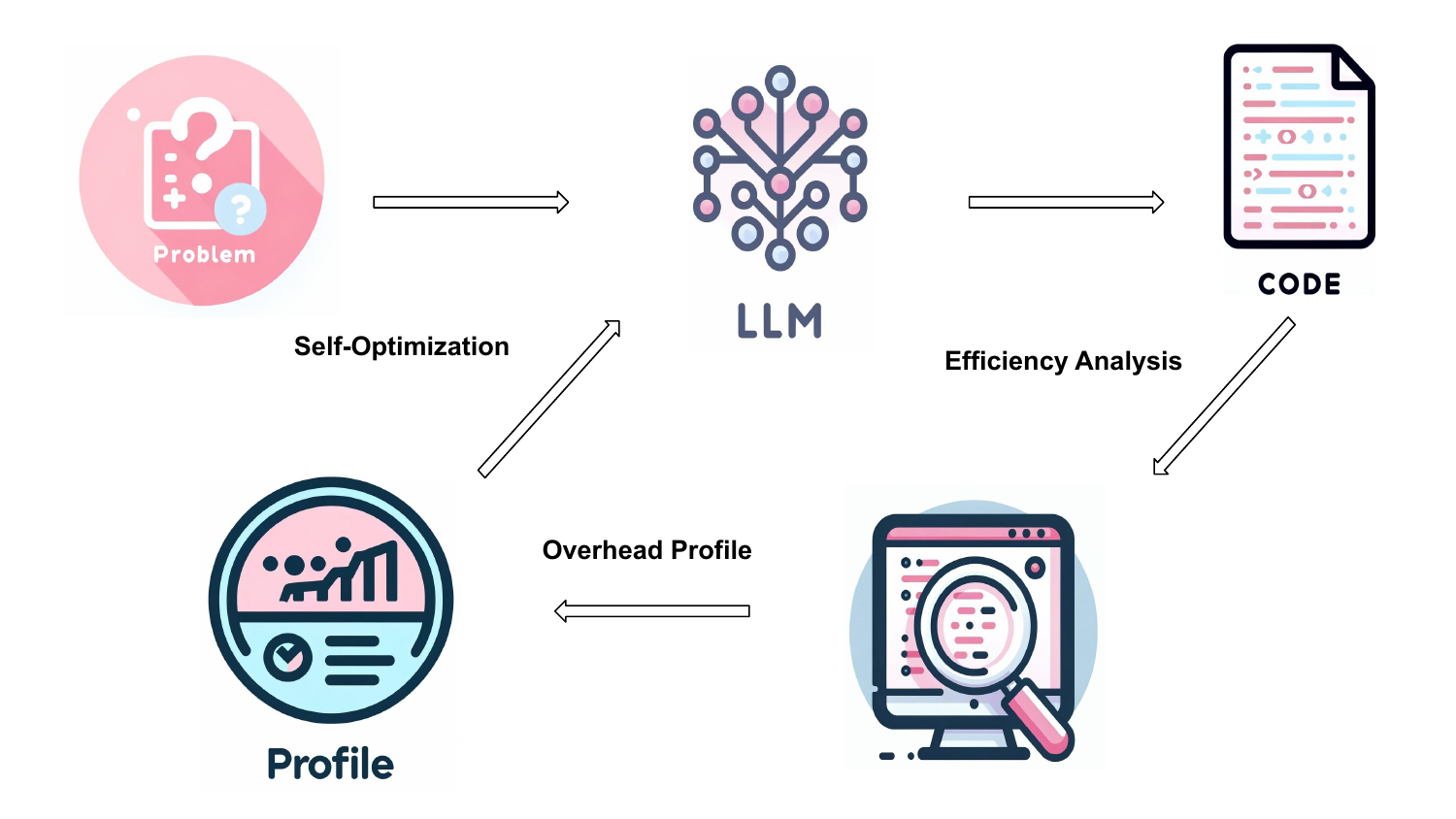}
    \vspace{-1mm}
    \caption{Pipeline of \name. LLMs first generate code for the given problem. This code is then executed locally to gather overhead profiles. These profiles are subsequently utilized by the LLMs to optimize the code in successive iterations, thereby enhancing the overall efficiency of the generated code. A comprehensive illustration is provided in the Appendix~\cref{fig:effective_case_task_description}-\cref{fig:effective_case_task_refine_code_memory}.} 
    \label{fig:framework}
    \vspace{-2mm}
\end{figure*}

Inspired by the optimization strategies employed by human coders on coding platforms, we propose a framework \name to enhance the efficiency of LLM-generated code. Human coders typically analyze execution time and memory usage profiles to identify bottlenecks and optimize their code. \name leverages this principle by integrating a self-optimization loop into the code generation process. As illustrated in Figure~\ref{fig:framework}, 
\name consists of three main components: \textbf{Code Generation}, \textbf{Overhead Profiling}, and \textbf{Code Refinement}, each playing a crucial role in the self-optimization process.

\subsection{Code Generation}
Given a task description or code generation requirement, the LLM generates an initial version of the code. The LLM takes the task description as input and produces code that aims to solve the task.

\subsection{Overhead Profiling}
The generated code is executed locally to capture its execution time and memory usage overhead profiles. During this step, the code is run on a set of open test cases, and the execution time and memory usage for each line of code are recorded. This information forms the overhead profiles that provide insights into the efficiency of the generated code.

\textbf{Execution Time Profiling \xspace} In this step, we measure the execution time of each line of code to identify potential bottlenecks and inefficiencies. To perform execution time profiling, we utilize the \texttt{line\_profiler} library in Python.
During the profiling process, we run the generated code on a set of open test cases provided by the dataset. The \texttt{line\_profiler} library tracks the execution time of each line of code for all the test cases combined. This helps us assess the code's performance under different conditions and identify any performance bottlenecks. The execution time profiling results are reported based on the total consumption for all open test cases. The profiling output includes information such as the line number, the number of times each line is executed, and the total time spent on each line. These profiles serve as input for the subsequent code refinement step.

\textbf{Memory Usage Profiling \xspace} Memory usage profiling is another essential aspect of the \name framework. It helps us understand how the generated code utilizes memory resources and identifies any memory-related inefficiencies or leaks. To profile memory usage, we employ the \texttt{memory\_profiler} library in Python.
During the memory usage profiling process, we run the generated code on the set of open test cases. The \texttt{memory\_profiler} library monitors the memory usage of each line of code throughout the execution of all the test cases combined. It captures the memory usage at different points, such as before and after function calls, loop iterations, and memory allocation statements. The memory usage profiling results are reported based on the total consumption for all open test cases. The profiling output includes information such as the line number and the corresponding memory usage. These profiles provide valuable insights into the memory efficiency of the generated code and help identify areas for optimization.

\subsection{Code Refinement}
This component leverages the execution time and memory usage profiles to optimize the generated code. 
In this step, the LLM analyzes the overhead profiles to refine the code for better efficiency. To enable self-optimization, we feed the overhead profiles back into the LLM along with the generated code. The LLM analyzes patterns in the overhead profiles, such as high execution time or excessive memory usage, and correlates them with specific code segments. It then applies optimization techniques, such as loop unrolling, memorization, data structure optimization, algorithm substitution, and code simplification, to improve the efficiency of the identified code segments.

During the self-optimization process, the LLM considers factors such as the impact of each optimization on the overall efficiency, the trade-offs between execution time and memory usage, and the preservation of code correctness. It aims to strike a balance between performance improvement and maintaining the functional integrity of the code. The LLM iteratively refines the code based on the overhead profiles, applying optimizations until a satisfactory level of efficiency is achieved or a predefined number of iterations is reached. The optimized code is then validated against the open test cases to ensure its functional correctness. By leveraging the execution time and memory usage profiles, the self-optimization step enables the LLM to improve the efficiency of the generated code.

\subsection{Prompt Construction}
We carefully design prompts to guide LLMs in optimizing code efficiency while ensuring the optimized code passes predefined test cases. The prompt template (\cref{fig:prompt_template}) used in \name's self-optimization stage includes a task description, test case, initial code, overhead analysis, and optimization rules. The task description provides context and requirements, the test case ensures correctness, and the initial code is the starting point for optimization. The overhead analysis highlights performance metrics and areas for improvement, while the optimization rules focus the LLM on enhancing efficiency, encapsulating the optimized code, and excluding the test case from the code block. This comprehensive prompt equips the LLM with the necessary information to effectively optimize code, maintain consistency across models and tasks, and facilitate comparison of their code optimization capabilities, advancing the field of LLM-driven code optimization. Details of the template can be found in Appendex~\ref{app:template}.

\section{Evaluation}
\label{evaluation}

\subsection{Dataset and Metrics}
We evaluate \name on EffiBench~\citep{huang2024effibench}. Following their settings, we use Execution Time (ET), Normalized Execution Time (NET), Max Memory Usage (MU), Normalized Max Memory Usage (NMU), Total Memory Usage (TMU), and Normalized Total Memory Usage (NTMU) as metrics. We provide a detailed definition of these metrics in~\ref{app:detailed-metric}. Following the setup of EffiBench, evaluation metrics were only calculated on the tasks that generated code for both the initial version and \name optimized code that can pass all private test cases provided by the dataset\footnote{Some LLMs may not generate correct initial code, we do not report the efficiency results for it.}.

Following \citet{huang2024effibench}, we utilize the open test cases to calculate the efficiency metrics during the self-optimization process, while private test cases provided by EffiBench were used for the final result evaluation. For HumanEval and MBPP datasets, we set the test cases provided by HumanEval and MBPP as open test cases, while test cases provided by EvalPlus \cite{liu2024your} (i.e., HumanEval-Plus, MBPP-Plus) as private test cases that were used to calculate the final results.

\subsection{Implementation Details}
All of the experiments are conducted in an edge server with 
an Intel Xeon Platinum 8336C CPU with 128 cores, and  8 * NVIDIA A100-SXM GPUs Total memory capacity of 2.0TiB.

\noindent\textbf{Models } 
We evaluate \name's effectiveness on both open-source and closed-source LLMs\footnote{We provide results for more LLMs in Table~\ref{tab:morellms} in Appendix.}. For open-source models, we evaluate \name with OpenCodeInterpreter (1.3B, 6.7B, and 33B) \cite{zheng2024opencodeinterpreter}, DeepSeek-Instruct (1.3B, 6.7B, and 33B) \cite{guo2024deepseek}, CodeLlama (7B, 13B, 34B, and 70B) \cite{Roziere2023}, XwinCoder (7B and 34B) \cite{xwin-lm}, StarCoder (3B, 7B, and 15B) \cite{LiStarCoder203}, and WizardCoder-13B \cite{luo2023wizardcoder}, where the detailed versions of LLMs are demonstrated in supplementary file.
For closed-source models, we use GPT-3.5-Turbo, GPT-4-Turbo, GPT-4~\citep{GPT4}, Claude-3-haiku, and Claude-3-sonnet\footnote{We do not include Claude-3-opus in our experiments due to limited resources.}.
These LLMs have achieved competitive pass@1 scores in various code generation tasks~\citep{ChenCodex2021, Austin2021, liu2024your}.

\noindent\textbf{Setup }
We first collect the generated code from each LLM and evaluate its correctness using open test cases. Only the code that passes all test cases is considered for efficiency evaluation. This approach ensures consistency in the evaluated tasks across different self-optimization iterations, as \name focuses on improving the efficiency of initially correct code without altering its pass@1 score. By evaluating a diverse set of open-source and closed-source LLMs, we aim to provide a comprehensive assessment of the efficiency of LLM-generated code and the effectiveness of \name in improving code efficiency across different models and architectures.

\begin{table*}\scriptsize
    \centering
    \caption{Code efficiency of LLMs with \name on EffiBench. The percentage in the brackets indicates the extent of the reduction for each respective item. Top performing LLMs are highlighted.}
    \label{tab:end2end}
    \begin{tabular}{lrrrrrr}
        \toprule
        Model&\textbf{ET (s)}&\textbf{NET}&\textbf{MU (Mb)}&\textbf{NMU}&\textbf{TMU (Mb*s)}&\textbf{NTMU}\\
        \midrule
\multirow{2}{*}{DeepSeek-6.7B-Ins}&0.37&2.60&259.73&7.25&555.18&67.70\\
&0.34 ({\color{problemcolor} 8.1\%})&2.37 ({\color{modelcolor} 8.8\%})&36.97 \textbf{({\color{problemcolor} 85.8\%})}&1.00 \textbf{({\color{modelcolor} 86.2\%})}&13.66 ({\color{problemcolor} 97.5\%})&1.46 ({\color{modelcolor} 97.8\%})\\
\multirow{2}{*}{CodeLlama-70b}&0.52&3.93&109.61&3.57&203.92&54.15\\
&0.47 ({\color{problemcolor} 9.6\%})&3.84 ({\color{modelcolor} 2.3\%})&26.42 ({\color{problemcolor} 75.9\%})&1.00 ({\color{modelcolor} 72.0\%})&14.53 ({\color{problemcolor} 92.9\%})&6.52 ({\color{modelcolor} 88.0\%})\\
\multirow{2}{*}{StarCoder2-15B}&0.93&7.58&26.35&1.00&22.02&10.88\\
&0.12 \textbf{({\color{problemcolor} 87.1\%})}&1.03 \textbf{({\color{modelcolor} 86.4\%})}&27.67 ({\color{problemcolor} -5.0\%})&1.01 ({\color{modelcolor} -1.0\%})&2.03 \textbf{({\color{problemcolor} 90.8\%})}&1.06 \textbf{({\color{modelcolor} 90.3\%})}\\

\multirow{2}{*}{GPT-3.5-Turbo-0301}&0.36&2.50&91.25&2.45&157.50&19.75\\
&0.28 \textbf{({\color{problemcolor} 22.2\%})}&2.01 \textbf{({\color{modelcolor} 19.6\%})}&36.08 \textbf{({\color{problemcolor} 60.5\%})}&0.99 \textbf{({\color{modelcolor} 59.6\%})}&12.43 \textbf{({\color{problemcolor} 92.1\%})}&1.64 \textbf{({\color{modelcolor} 91.7\%})}\\
\multirow{2}{*}{GPT-4}&0.31&2.19&80.88&2.26&129.91&17.90\\
&0.28 ({\color{problemcolor} 9.7\%})&2.06 ({\color{modelcolor} 5.9\%})&63.82 ({\color{problemcolor} 21.1\%})&1.83 ({\color{modelcolor} 19.0\%})&80.74 ({\color{problemcolor} 37.8\%})&11.86 ({\color{modelcolor} 33.7\%})\\
\multirow{2}{*}{Claude-3-Sonnet}&0.42&2.90&60.46&1.62&82.52&10.12\\
&0.35 ({\color{problemcolor} 16.7\%})&2.47 ({\color{modelcolor} 14.8\%})&42.31 ({\color{problemcolor} 30.0\%})&1.17 ({\color{modelcolor} 27.8\%})&28.95 ({\color{problemcolor} 64.9\%})&3.76 ({\color{modelcolor} 62.8\%})\\
        \bottomrule
    \end{tabular}
    \vspace{-0.3cm}
\end{table*}

\subsection{Main Results}

\textbf{Open-source LLMs}
As shown in \cref{tab:end2end}, we observe that the efficiency metrics for all models have been increased in most experiments once we apply \name to optimize the efficiency of LLM-generated code. For example, in OpenCodeInterpreter-1.3B, the execution time for its generated code decreases from 1.60 (s) to 1.29 (s), a reduction of 19.4\% in execution time. Additionally, the TMU of OpenCodeInterpreter-1.3B decreases from 89.16 (Mb*s) to 70.63 (Mb*s). Furthermore, in certain edge cases, \name significantly enhances efficiency. For example, the ET of StarCoder2-15B decreases from 0.93 (s) to 0.12 (s) and the NET also decreases from 7.48 to 1.03, reducing execution time requirements by 87.1\% compared to the initial code. The MU and NMU of DeepSeek-6.7B-Ins also decrease from 259.73 (Mb) and 7.25 to 36.97 (Mb) and 1.06, reducing the maximum memory consumption by 85.8\% for the code execution requirement. Moreover, we can also observe that the TMU and NTMU of StarCoder2-15B also decrease from 22.02 (Mb*s) and 10.88 to 2.03 (Mb*s) and 1.06, which decreases 90.8\% memory consumption during the execution process. These results demonstrate the effectiveness of \name in optimizing the code generated by open-source LLMs.

\textbf{Closed-source LLMs}
Similar to open-source LLMs, we observe that the efficiency metrics for most closed-source LLMs have been improved after applying \name to optimize the efficiency of the generated code. For instance, the execution time for code generated by GPT-3.5-Turbo-0301 decreases from 0.36 (s) to 0.28 (s), reducing the execution time by 22.2\%. The MU and NMU of GPT-3.5-Turbo-0301 also decrease from 91.25 (Mb) and 2.45 to 36.08 (Mb) and 0.99, respectively, which reduces the max memory consumption for code execution by 60.5\%. Furthermore, the TMU and NTMU of GPT-3.5-Turbo-0301 decrease from 157.50 (Mb*s) and 19.75 to 12.43 (Mb*s) and 1.64, respectively, decreasing memory consumption during the execution process by 92.1\%. 

The improvements in efficiency metrics across both open-source and closed-source LLMs highlight the generalizability and adaptability of \name. By iteratively refining the generated code based on efficiency profiler feedback, \name enables LLMs to produce more efficient code without compromising the correctness of the generated solutions. The consistent improvements across various models and architectures demonstrate the potential of \name as a model-agnostic approach for optimizing the efficiency of LLM-generated code in real-world applications.

\begin{table*}\scriptsize
    \centering
    \caption{Effect of the number of self-optimization steps in \name.}
    \begin{tabular}{lrrrrrr}
        \toprule
        Steps&\textbf{ET (s)}&\textbf{NET}&\textbf{MU (Mb)}&\textbf{NMU}&\textbf{TMU (Mb*s)}&\textbf{NTMU}\\
        \midrule
        \multicolumn{7}{l}{CodeLlama-70B}\\
        \midrule
0&0.52&3.93&109.61&3.57&203.92&54.15\\
1&0.48 ({\color{problemcolor} 7.7\%})&3.94 ({\color{modelcolor} -0.3\%})&26.47 ({\color{problemcolor} 75.9\%})&1.00 ({\color{modelcolor} 72.0\%})&14.91 ({\color{problemcolor} 92.7\%})&6.69 ({\color{modelcolor} 87.6\%})\\
2&0.48 ({\color{problemcolor} 7.7\%})&3.89 ({\color{modelcolor} 1.0\%})&26.47 ({\color{problemcolor} 75.9\%})&1.00 ({\color{modelcolor} 72.0\%})&14.69 ({\color{problemcolor} 92.8\%})&6.60 ({\color{modelcolor} 87.8\%})\\
3&0.47 ({\color{problemcolor} 9.6\%})&3.85 ({\color{modelcolor} 2.0\%})&26.42 ({\color{problemcolor} 75.9\%})&1.00 ({\color{modelcolor} 72.0\%})&14.60 ({\color{problemcolor} 92.8\%})&6.56 ({\color{modelcolor} 87.9\%})\\
4&0.47 ({\color{problemcolor} 9.6\%})&3.84 ({\color{modelcolor} 2.3\%})&26.42 ({\color{problemcolor} 75.9\%})&1.00 ({\color{modelcolor} 72.0\%})&14.54 ({\color{problemcolor} 92.9\%})&6.53 ({\color{modelcolor} 87.9\%})\\
5&0.47 ({\color{problemcolor} 9.6\%})&3.84 ({\color{modelcolor} 2.3\%})&26.42 ({\color{problemcolor} 75.9\%})&1.00 ({\color{modelcolor} 72.0\%})&14.53 ({\color{problemcolor} 92.9\%})&6.52 ({\color{modelcolor} 88.0\%})\\
        \midrule
        \multicolumn{7}{l}{GPT-3.5-Turbo-0301}\\
        \midrule
0& 0.36&2.50&91.25&2.45&157.50&19.75\\
1&0.33 ({\color{problemcolor} 8.3\%})&2.35 ({\color{modelcolor} 6.0\%})&36.09 ({\color{problemcolor} 60.4\%})&0.99 ({\color{modelcolor} 59.6\%})&13.70 ({\color{problemcolor} 91.3\%})&1.81 ({\color{modelcolor} 90.8\%})\\
2&0.31 ({\color{problemcolor} 13.9\%})&2.18 ({\color{modelcolor} 12.8\%})&36.09 ({\color{problemcolor} 60.4\%})&0.99 ({\color{modelcolor} 59.6\%})&13.04 ({\color{problemcolor} 91.7\%})&1.72 ({\color{modelcolor} 91.3\%})\\
3&0.29 ({\color{problemcolor} 19.4\%})&2.06 ({\color{modelcolor} 17.6\%})&36.08 ({\color{problemcolor} 60.5\%})&0.99 ({\color{modelcolor} 59.6\%})&12.57 ({\color{problemcolor} 92.0\%})&1.66 ({\color{modelcolor} 91.6\%})\\
4&0.29 ({\color{problemcolor} 19.4\%})&2.03 ({\color{modelcolor} 18.8\%})&36.08 ({\color{problemcolor} 60.5\%})&0.99 ({\color{modelcolor} 59.6\%})&12.50 ({\color{problemcolor} 92.1\%})&1.65 ({\color{modelcolor} 91.6\%})\\
5&0.28 ({\color{problemcolor} 22.2\%})&2.01 ({\color{modelcolor} 19.6\%})&36.08 ({\color{problemcolor} 60.5\%})&0.99 ({\color{modelcolor} 59.6\%})&12.43 ({\color{problemcolor} 92.1\%})&1.64 ({\color{modelcolor} 91.7\%})\\
        \bottomrule
    \end{tabular}
    \label{tab:steps}
\end{table*}

\subsection{Impact of Self-Optimization Steps}
To investigate the impact of the number of self-optimization steps on the efficiency of the \name-optimized code, we conduct an ablation study by varying the number of steps from 0 to 5. \cref{tab:steps} for CodeLlama-70B and GPT-3.5-Turbo-0301 at different self-optimization steps. 

\noindent \textbf{CodeLlama-70B} We can observe that after the first self-optimization step, the MU decreases from 109.61 (Mb) to 26.47 (Mb), reducing 75.9\% maximum memory requirement compared with the initial code generated by CodeLlama-70B. Similarly, the TMU decreases from 54.15 (Mb*s) to 6.69 (Mb*s), reducing 87.6\% of memory consumption during code execution. As the number of steps increases, the efficiency metrics gradually improve. By the fifth step, the ET reaches 0.47 (s), reducing the 1.9\% execution time requirement compared with the first-step generated code, and the TMU settles at 14.53, reducing 0.2\% total memory usage from the first step. 

\noindent \textbf{GPT-3.5-Turbo-0301} Similar to CodeLlama-70B, the MU decreases from 91.25 (Mb) to 36.09 (Mb) after the first self-optimization step, reducing 60.4\% maximum memory requirement compared with the initial code. The TMU also shows a substantial reduction from 157.50 (Mb*s) to 13.70 (Mb*s), reducing 91.3\% of memory consumption during code execution. As the number of steps increases, the efficiency metrics continue to improve steadily. By the fifth step, the ET reaches 0.28 (s), reducing the 15.2\% execution time requirement compared with the first-step generated code, and the TMU settles at 12.43 (Mb*s), reducing 9.3\% total memory usage from the first step.

\cref{tab:steps} demonstrates the significant impact of the number of self-optimization steps on the efficiency of the \name-optimized code. For both CodeLlama-70B and GPT-3.5-Turbo-0301, the first self-optimization step yields the most substantial improvements in code efficiency. 
As the number of steps increases, the efficiency metrics continue to improve, albeit with diminishing returns. By the fifth step, the efficiency metrics reach their lowest values, demonstrating the effectiveness of \name's iterative self-optimization approach in enhancing the efficiency of LLM-generated code. The evaluation results highlight that the majority of efficiency improvements occur in the first few steps, with subsequent steps contributing to further refinements of the optimized code.

\begin{table*}
\tiny
    \centering
    \caption{Contribution of different components in \name. We evaluate how different feedback profilers affect the efficiency of LLM-generated code. Unsupervised self-refine only requires LLMs to optimize the efficiency of the code. Result-Aware Self-Refine feedback the ET, MU, and TMU to the LLMs and require it to improve the efficiency. 
    Memory Profiler and Execution Time Profiler feedback the memory profiler and execution time profiler to the LLMs and then LLMs can based on the profile optimize the efficiency of the code.
    }
    \begin{tabular}{lrrrrrr}
        \toprule
        Optimization Profile&\textbf{ET (s)}&\textbf{NET}&\textbf{MU (Mb)}&\textbf{NMU}&\textbf{TMU (Mb*s)}&\textbf{NTMU}\\
        \midrule
        \multicolumn{7}{l}{CodeLlama-70B}\\
        \midrule
Initial Version&0.52&3.93&109.61&3.57&203.92&54.15\\
Unsupervised Self-Refine&0.79 ({\color{problemcolor} -51.9\%})&6.87 ({\color{modelcolor} -74.8\%})&279.41 ({\color{problemcolor} -154.9\%})&10.58 ({\color{modelcolor} -196.4\%})&1261.83 ({\color{problemcolor} -518.8\%})&600.95 ({\color{modelcolor} -1009.8\%})\\
Result-Aware Self-Refine&0.79 ({\color{problemcolor} -51.9\%})&6.87 ({\color{modelcolor} -74.8\%})&282.57 ({\color{problemcolor} -157.8\%})&10.70 ({\color{modelcolor} -199.7\%})&1270.93 ({\color{problemcolor} -523.2\%})&605.29 ({\color{modelcolor} -1017.8\%})\\
Memory Profiler&0.53 ({\color{problemcolor} -1.9\%})&4.34 ({\color{modelcolor} -10.4\%})&26.38 ({\color{problemcolor} 75.9\%})&0.99 ({\color{modelcolor} 72.3\%})&15.77 ({\color{problemcolor} 92.3\%})&7.06 ({\color{modelcolor} 87.0\%})\\
Execution Time Profiler&0.51 ({\color{problemcolor} 1.9\%})&4.17 ({\color{modelcolor} -6.1\%})&26.44 ({\color{problemcolor} 75.9\%})&1.00 ({\color{modelcolor} 72.0\%})&15.53 ({\color{problemcolor} 92.4\%})&6.97 ({\color{modelcolor} 87.1\%})\\
\name&0.47 ({\color{problemcolor} 9.6\%})&3.84 ({\color{modelcolor} 2.3\%})&26.42 ({\color{problemcolor} 75.9\%})&1.00 ({\color{modelcolor} 72.0\%})&14.53 ({\color{problemcolor} 92.9\%})&6.52 ({\color{modelcolor} 88.0\%})\\
        \midrule
        \multicolumn{7}{l}{GPT-3.5-Turbo-0301}\\
        \midrule

Initial Version& 0.36&2.50&91.25&2.45&157.50&19.75\\
Unsupervised Self-Refine&0.32 ({\color{problemcolor} 11.1\%})&2.46 ({\color{modelcolor} 1.6\%})&78.39 ({\color{problemcolor} 14.1\%})&2.12 ({\color{modelcolor} 13.5\%})&312.99 ({\color{problemcolor} -98.7\%})&42.42 ({\color{modelcolor} -114.8\%})\\
Result-Aware Self-Refine&0.30 ({\color{problemcolor} 16.7\%})&2.25 ({\color{modelcolor} 10.0\%})&58.65 ({\color{problemcolor} 35.7\%})&1.61 ({\color{modelcolor} 34.3\%})&195.49 ({\color{problemcolor} -24.1\%})&27.16 ({\color{modelcolor} -37.5\%})\\

Memory Profiler&0.34 ({\color{problemcolor} 5.6\%})&2.40 ({\color{modelcolor} 4.0\%})&36.85 ({\color{problemcolor} 59.6\%})&1.00 ({\color{modelcolor} 59.2\%})&16.34 ({\color{problemcolor} 89.6\%})&2.10 ({\color{modelcolor} 89.4\%})\\
Execution Time Profiler&0.33 ({\color{problemcolor} 8.3\%})&2.34 ({\color{modelcolor} 6.4\%})&36.43 ({\color{problemcolor} 60.1\%})&0.99 ({\color{modelcolor} 59.6\%})&14.07 ({\color{problemcolor} 91.1\%})&1.81 ({\color{modelcolor} 90.8\%})\\
\name&0.28 ({\color{problemcolor} 22.2\%})&2.01 ({\color{modelcolor} 19.6\%})&36.08 ({\color{problemcolor} 60.5\%})&0.99 ({\color{modelcolor} 59.6\%})&12.43 ({\color{problemcolor} 92.1\%})&1.64 ({\color{modelcolor} 91.7\%})\\
        \bottomrule
    \end{tabular}
    \label{tab:component}
    \vspace{-0.3cm}
\end{table*}

\subsection{Feedback of Overhead Profile}

To show the effectiveness of the overhead profile in guiding LLMs to refine their generated code, we compare the performance of \name with two alternative approaches: Unsupervised Self-Refine and Result-Aware Self-Refine~\citep{MadaanSelfRefine23, ShinnReflexion23}. Unsupervised Self-Refine uses a prompt that directly requires the LLM to refine the code without providing additional information. Result-Aware Self-Refine feeds the ET, MU, and TMU, then requires the LLM to refine the code based on these metrics. \cref{tab:component} presents the code efficiency metrics for CodeLlama-70B and GPT-3.5-Turbo-0301 using different code refinement approaches. 

\noindent \textbf{CodeLlama-70B} Unsupervised Self-Refine and Result-Aware Self-Refine result in significant increases in ET, memory usage (MU), and TMU compared to the initial version. Unsupervised Self-Refine increases ET by 51.9\%, MU by 154.9\%, and TMU by 518.8\%, while Result-Aware Self-Refine increases ET by 51.9\%, MU by 157.8\%, and TMU by 523.2\%. In contrast, \name incorporates the overhead profile feedback and achieves a 9.6\% reduction in ET, a 75.9\% reduction in MU, and a 92.9\% reduction in TMU compared to the initial version.

\noindent \textbf{GPT-3.5-Turbo-0301} Unsupervised Self-Refine and Result-Aware Self-Refine show some improvements in ET and MU compared to the initial version. Unsupervised Self-Refine reduces ET by 11.1\% and MU by 14.1\%, while Result-Aware Self-Refine reduces ET by 16.7\% and MU by 35.7\%. However, both approaches lead to substantial increases in TMU, with Unsupervised Self-Refine increasing TMU by 98.7\% and Result-Aware Self-Refine increasing TMU by 24.1\%. On the other hand, \name achieves a 22.2\% reduction in ET, a 60.5\% reduction in MU, and a 92.1\% reduction in TMU compared to the initial version.

These results highlight the importance of the overhead profile feedback in guiding LLMs to generate more efficient code. Without the overhead profile, the code refinement process using alternative prompts fails to improve code efficiency and even leads to significant performance degradation. The overhead profile provides valuable insights into the resource consumption of the generated code, enabling LLMs to make targeted optimizations and achieve substantial efficiency improvements.

\subsection{Discussion}
\label{discussion}

\begin{table*}\scriptsize
    \centering
    \caption{Results of \name on HumanEval dataset, where we evaluate CodeLlama family generated code's efficiency. Full results are listed in Appendix \cref{tab:humaneval}.}
    \begin{tabular}{lrrrrrr}
        \toprule
        Steps&\textbf{ET (s)}&\textbf{NET}&\textbf{MU (Mb)}&\textbf{NMU}&\textbf{TMU (Mb*s)}&\textbf{NTMU}\\
        \midrule
CodeLlama-7b&0.20&0.71&57.39&0.91&7.08&0.70\\
&0.18 ({\color{problemcolor} 10.0\%})&0.63 ({\color{modelcolor} 11.3\%})&57.07 ({\color{problemcolor} 0.6\%})&0.91 ({\color{modelcolor} 0.0\%})&6.18 ({\color{problemcolor} 12.7\%})&0.61 ({\color{modelcolor} 12.9\%})\\
CodeLlama-13b&0.23&0.95&58.13&0.96&7.97&0.94\\
&0.20 ({\color{problemcolor} 13.0\%})&0.80 ({\color{modelcolor} 15.8\%})&58.03 ({\color{problemcolor} 0.2\%})&0.96 ({\color{modelcolor} 0.0\%})&6.64 ({\color{problemcolor} 16.7\%})&0.79 ({\color{modelcolor} 16.0\%})\\
CodeLlama-34b&0.24&0.95&61.79&1.01&8.45&0.96\\
&0.21 ({\color{problemcolor} 12.5\%})&0.81 ({\color{modelcolor} 14.7\%})&61.55 ({\color{problemcolor} 0.4\%})&1.00 ({\color{modelcolor} 1.0\%})&6.99 ({\color{problemcolor} 17.3\%})&0.80 ({\color{modelcolor} 16.7\%})\\
CodeLlama-70b&0.21&0.93&60.19&1.01&6.76&1.01\\
&0.18 ({\color{problemcolor} 14.3\%})&0.79 ({\color{modelcolor} 15.1\%})&59.49 ({\color{problemcolor} 1.2\%})&1.00 ({\color{modelcolor} 1.0\%})&5.75 ({\color{problemcolor} 14.9\%})&0.86 ({\color{modelcolor} 14.9\%})\\
        \bottomrule
    \end{tabular}
    \vspace{-0.3cm}
    \label{tab:humaneval_small}
\end{table*}

\begin{table*}[ht!]\scriptsize
    \centering
    \caption{Code efficiency of widely-studied LLMs reported by \name.}
    \begin{tabular}{lrrrrrr}
        \toprule
        Model&\textbf{ET (s)}&\textbf{NET}&\textbf{MU (Mb)}&\textbf{NMU}&\textbf{TMU (Mb*s)}&\textbf{NTMU}\\
        \midrule
\multirow{2}{*}{Phind-CodeLlama-34B-v2}&0.52&3.28&157.16&3.36&337.30&24.44\\
&0.40 ({\color{problemcolor} 23.1\%})&2.51 ({\color{modelcolor} 23.5\%})&68.27 ({\color{problemcolor} 56.6\%})&1.45 ({\color{modelcolor} 56.8\%})&65.64 ({\color{problemcolor} 80.5\%})&4.86 ({\color{modelcolor} 80.1\%})\\
\multirow{2}{*}{Artigenz-Coder-DS-6.7B}&0.39&2.75&65.73&1.70&95.65&10.87\\
&0.32 ({\color{problemcolor} 17.9\%})&2.30 ({\color{modelcolor} 16.4\%})&59.00 ({\color{problemcolor} 10.2\%})&1.62 ({\color{modelcolor} 4.7\%})&79.67 ({\color{problemcolor} 16.7\%})&10.73 ({\color{modelcolor} 1.3\%})\\
\multirow{2}{*}{Magicoder-S-DS-6.7B}&0.22&1.59&40.19&1.09&17.58&2.28\\
&0.21 ({\color{problemcolor} 4.5\%})&1.50 ({\color{modelcolor} 5.7\%})&38.29 ({\color{problemcolor} 4.7\%})&1.07 ({\color{modelcolor} 1.8\%})&15.27 ({\color{problemcolor} 13.1\%})&2.22 ({\color{modelcolor} 2.6\%})\\
\multirow{2}{*}{Mistral-7B-codealpaca-lora}&2.36&18.40&28.88&1.00&57.92&24.36\\
&1.45 ({\color{problemcolor} 38.6\%})&12.17 ({\color{modelcolor} 33.9\%})&27.45 ({\color{problemcolor} 5.0\%})&1.03 ({\color{modelcolor} -3.0\%})&35.46 ({\color{problemcolor} 38.8\%})&17.28 ({\color{modelcolor} 29.1\%})\\
CodeFuse-DeepSeek-33B&0.40&3.10&70.39&2.06&191.15&32.20\\
&0.39 ({\color{problemcolor} 2.5\%})&3.01 ({\color{modelcolor} 2.9\%})&63.22 ({\color{problemcolor} 10.2\%})&1.85 ({\color{modelcolor} 10.2\%})&156.81 ({\color{problemcolor} 18.0\%})&26.42 ({\color{modelcolor} 18.0\%})\\
CodeLlama-34b-hf&2.08&15.68&46.41&1.26&128.46&17.87\\
&1.95 ({\color{problemcolor} 6.3\%})&14.67 ({\color{modelcolor} 6.4\%})&46.40 ({\color{problemcolor} 0.0\%})&1.26 ({\color{modelcolor} 0.0\%})&125.22 ({\color{problemcolor} 2.5\%})&17.42 ({\color{modelcolor} 2.5\%})\\
speechless-starcoder2-15b&0.19&1.74&27.39&0.99&3.20&1.75\\
&0.13 ({\color{problemcolor} 31.6\%})&1.19 ({\color{modelcolor} 31.6\%})&27.25 ({\color{problemcolor} 0.5\%})&0.99 ({\color{modelcolor} 0.0\%})&2.17 ({\color{problemcolor} 32.2\%})&1.19 ({\color{modelcolor} 32.0\%})\\
gpt-3.5-turbo-0613&0.56&4.32&35.48&1.00&20.11&3.00\\
&0.49 ({\color{problemcolor} 12.5\%})&3.75 ({\color{modelcolor} 13.2\%})&35.47 ({\color{problemcolor} 0.0\%})&1.00 ({\color{modelcolor} 0.0\%})&17.84 ({\color{problemcolor} 11.3\%})&2.66 ({\color{modelcolor} 11.3\%})\\
        \bottomrule
    \end{tabular}
    \label{tab:morellms}
\end{table*}

\begin{table}
    \centering
    \caption{Pass@1 of LLMs generated initial code and \name optimized code.}
    \begin{tabular}{l|rr}
    \toprule
         Model&Initial Pass@1&\name Pass@1\\
         \bottomrule
OpenCodeInterpreter-DS-1.3B&5.8&5.4\\
OpenCodeInterpreter-DS-6.7B&13.6&13.2\\
OpenCodeInterpreter-DS-33B&24.7&24.4\\
deepseek-1.3b-Ins&4.8&4.5\\
deepseek-6.7b-Ins&7.2&7.0\\
deepseek-33b-Ins&10.0&9.9\\
CodeLlama-7b&7.0&7.0\\
CodeLlama-13b&9.7&9.6\\
CodeLlama-34b&13.5&13.0\\
CodeLlama-70b&7.8&7.4\\
XwinCoder-13B&10.5&10.2\\
XwinCoder-34B&21.2&21.2\\
starcoder2-3b&1.6&1.2\\
starcoder2-7b&1.9&1.8\\
starcoder2-15b&0.7&0.4\\
WizardCoder-13B&4.0&3.9\\
\bottomrule
    \end{tabular}
    \label{tab:pass@1}
\end{table}

\noindent\textbf{Generalizability across benchmarks } In \cref{tab:end2end}, we evaluated \name's effectiveness on the EffiBench dataset. To illustrate \name's generalizability in other datasets, we conduct experiments on the HumanEval and MBPP datasets in Appendix \cref{tab:humaneval} and \cref{tab:mbpp}. We also provide \name's effectiveness on the HumanEval dataset in CodeLlama models in \cref{tab:humaneval_small}. We can observe that the coding efficiency of CodeLlama and other LLMs (See \cref{tab:humaneval}) also increases when we utilize \name to optimize LLM-generated code. For example, the ET of CodeLlama-70B decreases from 0.21 (s) to 0.18 (s), which reduces 14.3\% execution time. As shown in \cref{tab:humaneval} and \cref{tab:mbpp}, results demonstrate that \name can consistently improve the efficiency of LLM-generated code for other datasets.

\noindent\textbf{Generalizability across LLMs } In \cref{tab:end2end}, we evaluate \name's effectiveness on six types of open-source LLMs. To illustrate \name's generalizability in other LLMs, we also conduct experiments on other LLMs in~\cref{tab:morellms}. Our evaluation results demonstrate that \name can improve the efficiency of LLM-generated code for different LLMs. For example, the execution time of Mistral-7B-codealpaca-lora decreases from 2.36 (s) to 1.45 (s), which reduces 38.6\% execution time compared with the initial code. The total memory usage of Phind-CodeLlama-34B-v2 also decreases from 337.30 (Mb*s) to 65.64 (Mb*s), which reduces 80.5\% total memory requirement.

\noindent\textbf{Impact on correctness } We provide the pass@1 of LLM-generated initial code and \name optimized code for EffiBench in \cref{tab:pass@1}. We observe that the pass@1 of \name optimized code may be lower than LLM-generated initial code. The key reason is that during the self-optimization process, \name only uses public test cases to guide code efficiency optimization for correct initial code. However, since public test cases may not cover all edge cases in the private test cases (test cases used to evaluate pass@1 of LLMs), this can cause the pass@1 of \name generated code to be lower than the initial code. Nevertheless, we observe that the pass@1 of \name only decreases by about 0\% to 0.5\%, which means that only a few of the codes will be incorrect. As shown in \cref{tab:end2end}, we can observe that the code efficiency is largely increased. We believe that this minor decrease in pass@1 is worthwhile considering the significant efficiency gains.

\noindent\textbf{Case study }
To illustrate how \name improves the efficiency of LLM-generated code, we provide a case illustration in Appendix \cref{fig:effective_case_task_description}-\cref{fig:effective_case_task_refine_code_memory}. As shown in \cref{fig:effective_case_task_time}, we can observe that the execution time of the initial code is 23.59 (s) while in the self-optimized code, the execution time decreases from 23.59 (s) to 3.36 (s). The key reason is that in the initial code, the algorithm uses a standard unidirectional Breadth-First Search (BFS), which explores all possible states level by level starting from the initial state. This method results in a large number of states to explore, leading to significant computational overhead. In contrast, the self-optimized code employs a bidirectional BFS, which simultaneously searches from both the initial state and the target state. This reduces the search space by meeting in the middle, significantly decreasing the number of states that need to be explored and thereby improving the execution time.

\noindent\textbf{Error Analysis }
We also provide a case illustration to explain why some code efficiency does not improve significantly when \name is applied to LLM-generated code. As shown in Appendix \cref{fig:ineffective_case_task_description}-\cref{fig:ineffective_case_task_code_refine_code_memory}, we observe that the initial code only requires 0.0012 (s) to execute, while in the optimized code, the execution time is still 0.0011 (s). The key reason for this minimal improvement is that both implementations already operate with the same theoretical time complexity of $O(\log(\min(m, n)))$. Given the problem's constraints and small input sizes, the actual runtime differences are overshadowed by the inherent efficiency of the binary search algorithm. Additionally, the overhead of function calls and Python runtime operations can further minimize the observed performance gains. Therefore, while the optimized code may offer clearer partition management and slight improvements, the overall efficiency remains largely unchanged due to the already optimized nature of the initial approach.

\section{Conclusion}
\label{conclusion}

This paper focuses on the critical issue of efficiency in code generated by LLMs. While LLMs have shown impressive capabilities in code generation, their output often suffers from suboptimal efficiency, leading to slower execution and higher resource consumption. To tackle this challenge, we propose \name, a novel self-optimization framework that leverages execution overhead profiles to guide LLMs in improving code efficiency. Extensive experiments and analysis demonstrate that \name significantly enhances the efficiency of LLM-generated code, achieving substantial reductions in execution time and memory usage. For future work, we would like to investigate the application of \name to other programming tasks and languages, as well as explore the potential benefits of incorporating domain-specific knowledge into the optimization process. 

\section{ACKNOWLEDGMENT}

The work is supported in part by National Key R\&D Program of China (2022ZD0160201), HK RGC RIF (R7030-22), HK ITF (GHP/169/20SZ), a Huawei Flagship Research Grant in 2023, HK RGC GRF (Ref: 17208223 \& 17204424), and the HKU-CAS Joint Laboratory for Intelligent System Software.

\bibliography{neurips_2024}

\begin{thebibliography}{75}
\providecommand{\natexlab}[1]{#1}
\providecommand{\url}[1]{\texttt{#1}}
\expandafter\ifx\csname urlstyle\endcsname\relax
  \providecommand{\doi}[1]{doi: #1}\else
  \providecommand{\doi}{doi: \begingroup \urlstyle{rm}\Url}\fi

\bibitem[Ahmad et~al.(2023)Ahmad, Tushar, Chakraborty, and Chang]{AhmadTCC23}
Ahmad, W.~U., Tushar, M. G.~R., Chakraborty, S., and Chang, K.
\newblock {AVATAR:} {A} parallel corpus for java-python program translation.
\newblock In Rogers, A., Boyd{-}Graber, J.~L., and Okazaki, N. (eds.), \emph{Findings of the Association for Computational Linguistics: {ACL} 2023, Toronto, Canada, July 9-14, 2023}, pp.\  2268--2281. Association for Computational Linguistics, 2023.
\newblock \doi{10.18653/V1/2023.FINDINGS-ACL.143}.
\newblock URL \url{https://doi.org/10.18653/v1/2023.findings-acl.143}.

\bibitem[Ahmed \& Devanbu(2022)Ahmed and Devanbu]{AhmedD22}
Ahmed, T. and Devanbu, P.~T.
\newblock Few-shot training llms for project-specific code-summarization.
\newblock In \emph{37th {IEEE/ACM} International Conference on Automated Software Engineering, {ASE} 2022, Rochester, MI, USA, October 10-14, 2022}, pp.\  177:1--177:5. {ACM}, 2022.
\newblock \doi{10.1145/3551349.3559555}.
\newblock URL \url{https://doi.org/10.1145/3551349.3559555}.

\bibitem[Allal et~al.(2023)Allal, Li, Kocetkov, Mou, Akiki, Ferrandis, Muennighoff, Mishra, Gu, Dey, Umapathi, Anderson, Zi, Lamy{-}Poirier, Schoelkopf, Troshin, Abulkhanov, Romero, Lappert, Toni, del R{\'{\i}}o, Liu, Bose, Bhattacharyya, Zhuo, Yu, Villegas, Zocca, Mangrulkar, Lansky, Nguyen, Contractor, Villa, Li, Bahdanau, Jernite, Hughes, Fried, Guha, de~Vries, and von Werra]{Loubnasanta2023}
Allal, L.~B., Li, R., Kocetkov, D., Mou, C., Akiki, C., Ferrandis, C.~M., Muennighoff, N., Mishra, M., Gu, A., Dey, M., Umapathi, L.~K., Anderson, C.~J., Zi, Y., Lamy{-}Poirier, J., Schoelkopf, H., Troshin, S., Abulkhanov, D., Romero, M., Lappert, M., Toni, F.~D., del R{\'{\i}}o, B.~G., Liu, Q., Bose, S., Bhattacharyya, U., Zhuo, T.~Y., Yu, I., Villegas, P., Zocca, M., Mangrulkar, S., Lansky, D., Nguyen, H., Contractor, D., Villa, L., Li, J., Bahdanau, D., Jernite, Y., Hughes, S., Fried, D., Guha, A., de~Vries, H., and von Werra, L.
\newblock Santacoder: don't reach for the stars!
\newblock \emph{CoRR}, abs/2301.03988, 2023.
\newblock \doi{10.48550/ARXIV.2301.03988}.
\newblock URL \url{https://doi.org/10.48550/arXiv.2301.03988}.

\bibitem[Anil et~al.(2023)Anil, Borgeaud, Wu, Alayrac, Yu, Soricut, Schalkwyk, Dai, Hauth, Millican, Silver, Petrov, Johnson, Antonoglou, Schrittwieser, Glaese, Chen, Pitler, Lillicrap, Lazaridou, Firat, Molloy, Isard, Barham, Hennigan, Lee, Viola, Reynolds, Xu, Doherty, Collins, Meyer, Rutherford, Moreira, Ayoub, Goel, Tucker, Piqueras, Krikun, Barr, Savinov, Danihelka, Roelofs, White, Andreassen, von Glehn, Yagati, Kazemi, Gonzalez, Khalman, Sygnowski, and et~al.]{Gemini23}
Anil, R., Borgeaud, S., Wu, Y., Alayrac, J., Yu, J., Soricut, R., Schalkwyk, J., Dai, A.~M., Hauth, A., Millican, K., Silver, D., Petrov, S., Johnson, M., Antonoglou, I., Schrittwieser, J., Glaese, A., Chen, J., Pitler, E., Lillicrap, T.~P., Lazaridou, A., Firat, O., Molloy, J., Isard, M., Barham, P.~R., Hennigan, T., Lee, B., Viola, F., Reynolds, M., Xu, Y., Doherty, R., Collins, E., Meyer, C., Rutherford, E., Moreira, E., Ayoub, K., Goel, M., Tucker, G., Piqueras, E., Krikun, M., Barr, I., Savinov, N., Danihelka, I., Roelofs, B., White, A., Andreassen, A., von Glehn, T., Yagati, L., Kazemi, M., Gonzalez, L., Khalman, M., Sygnowski, J., and et~al.
\newblock Gemini: {A} family of highly capable multimodal models.
\newblock \emph{CoRR}, abs/2312.11805, 2023.
\newblock \doi{10.48550/ARXIV.2312.11805}.
\newblock URL \url{https://doi.org/10.48550/arXiv.2312.11805}.

\bibitem[Anthropic(2024)]{Claude3}
Anthropic.
\newblock Introducing the next generation of claude, 2024.
\newblock URL \url{https://www.anthropic.com/news/claude-3-family}.

\bibitem[Austin et~al.(2021)Austin, Odena, Nye, Bosma, Michalewski, Dohan, Jiang, Cai, Terry, Le, and Sutton]{Austin2021}
Austin, J., Odena, A., Nye, M.~I., Bosma, M., Michalewski, H., Dohan, D., Jiang, E., Cai, C.~J., Terry, M., Le, Q.~V., and Sutton, C.
\newblock Program synthesis with large language models.
\newblock \emph{CoRR}, abs/2108.07732, 2021.
\newblock URL \url{https://arxiv.org/abs/2108.07732}.

\bibitem[Boyd \& Fales(1983)Boyd and Fales]{boyd1983reflective}
Boyd, E.~M. and Fales, A.~W.
\newblock Reflective learning: Key to learning from experience.
\newblock \emph{Journal of humanistic psychology}, 23\penalty0 (2):\penalty0 99--117, 1983.

\bibitem[Brown et~al.(2020)Brown, Mann, Ryder, Subbiah, Kaplan, Dhariwal, Neelakantan, Shyam, Sastry, Askell, Agarwal, Herbert{-}Voss, Krueger, Henighan, Child, Ramesh, Ziegler, Wu, Winter, Hesse, Chen, Sigler, Litwin, Gray, Chess, Clark, Berner, McCandlish, Radford, Sutskever, and Amodei]{BrownMRSKDNSSAA20}
Brown, T.~B., Mann, B., Ryder, N., Subbiah, M., Kaplan, J., Dhariwal, P., Neelakantan, A., Shyam, P., Sastry, G., Askell, A., Agarwal, S., Herbert{-}Voss, A., Krueger, G., Henighan, T., Child, R., Ramesh, A., Ziegler, D.~M., Wu, J., Winter, C., Hesse, C., Chen, M., Sigler, E., Litwin, M., Gray, S., Chess, B., Clark, J., Berner, C., McCandlish, S., Radford, A., Sutskever, I., and Amodei, D.
\newblock Language models are few-shot learners.
\newblock In Larochelle, H., Ranzato, M., Hadsell, R., Balcan, M., and Lin, H. (eds.), \emph{Advances in Neural Information Processing Systems 33: Annual Conference on Neural Information Processing Systems 2020, NeurIPS 2020, December 6-12, 2020, virtual}, 2020.
\newblock URL \url{https://proceedings.neurips.cc/paper/2020/hash/1457c0d6bfcb4967418bfb8ac142f64a-Abstract.html}.

\bibitem[Capra et~al.(2012)Capra, Francalanci, and Slaughter]{capra2012software}
Capra, E., Francalanci, C., and Slaughter, S.~A.
\newblock Is software “green”? application development environments and energy efficiency in open source applications.
\newblock \emph{Information and Software Technology}, 54\penalty0 (1):\penalty0 60--71, 2012.

\bibitem[Chaudhary(2023)]{Codealpaca}
Chaudhary, S.
\newblock Code alpaca: An instruction-following llama model for code generation.
\newblock \url{https://github.com/sahil280114/codealpaca}, 2023.

\bibitem[Chen et~al.(2021)Chen, Tworek, Jun, Yuan, de~Oliveira~Pinto, Kaplan, Edwards, Burda, Joseph, Brockman, Ray, Puri, Krueger, Petrov, Khlaaf, Sastry, Mishkin, Chan, Gray, Ryder, Pavlov, Power, Kaiser, Bavarian, Winter, Tillet, Such, Cummings, Plappert, Chantzis, Barnes, Herbert{-}Voss, Guss, Nichol, Paino, Tezak, Tang, Babuschkin, Balaji, Jain, Saunders, Hesse, Carr, Leike, Achiam, Misra, Morikawa, Radford, Knight, Brundage, Murati, Mayer, Welinder, McGrew, Amodei, McCandlish, Sutskever, and Zaremba]{ChenCodex2021}
Chen, M., Tworek, J., Jun, H., Yuan, Q., de~Oliveira~Pinto, H.~P., Kaplan, J., Edwards, H., Burda, Y., Joseph, N., Brockman, G., Ray, A., Puri, R., Krueger, G., Petrov, M., Khlaaf, H., Sastry, G., Mishkin, P., Chan, B., Gray, S., Ryder, N., Pavlov, M., Power, A., Kaiser, L., Bavarian, M., Winter, C., Tillet, P., Such, F.~P., Cummings, D., Plappert, M., Chantzis, F., Barnes, E., Herbert{-}Voss, A., Guss, W.~H., Nichol, A., Paino, A., Tezak, N., Tang, J., Babuschkin, I., Balaji, S., Jain, S., Saunders, W., Hesse, C., Carr, A.~N., Leike, J., Achiam, J., Misra, V., Morikawa, E., Radford, A., Knight, M., Brundage, M., Murati, M., Mayer, K., Welinder, P., McGrew, B., Amodei, D., McCandlish, S., Sutskever, I., and Zaremba, W.
\newblock Evaluating large language models trained on code.
\newblock \emph{CoRR}, abs/2107.03374, 2021.
\newblock URL \url{https://arxiv.org/abs/2107.03374}.

\bibitem[Chen et~al.(2023)Chen, Lin, Sch{\"{a}}rli, and Zhou]{ChenSelfDebug23}
Chen, X., Lin, M., Sch{\"{a}}rli, N., and Zhou, D.
\newblock Teaching large language models to self-debug.
\newblock \emph{CoRR}, abs/2304.05128, 2023.
\newblock \doi{10.48550/ARXIV.2304.05128}.
\newblock URL \url{https://doi.org/10.48550/arXiv.2304.05128}.

\bibitem[Coignion et~al.(2024)Coignion, Quinton, and Rouvoy]{coignion2024performance}
Coignion, T., Quinton, C., and Rouvoy, R.
\newblock åå.
\newblock In \emph{Proceedings of the 28th International Conference on Evaluation and Assessment in Software Engineering}, pp.\  79--89, 2024.

\bibitem[Dai et~al.(2024)Dai, Lu, Feng, Ruan, Cheng, Tan, and Guo]{Dai2024mhpp}
Dai, J., Lu, J., Feng, Y., Ruan, R., Cheng, M., Tan, H., and Guo, Z.
\newblock {MHPP:} exploring the capabilities and limitations of language models beyond basic code generation.
\newblock \emph{CoRR}, abs/2405.11430, 2024.
\newblock \doi{10.48550/ARXIV.2405.11430}.
\newblock URL \url{https://doi.org/10.48550/arXiv.2405.11430}.

\bibitem[DeepSeekAI(2023)]{deepseekcoder}
DeepSeekAI.
\newblock Deepseek coder: Let the code write itself, 2023.
\newblock URL \url{https://deepseekcoder.github.io/}.

\bibitem[Deng et~al.(2023)Deng, Xia, Yang, Zhang, Yang, and Zhang]{Deng2023}
Deng, Y., Xia, C.~S., Yang, C., Zhang, S.~D., Yang, S., and Zhang, L.
\newblock Large language models are edge-case fuzzers: Testing deep learning libraries via fuzzgpt.
\newblock \emph{CoRR}, abs/2304.02014, 2023.
\newblock \doi{10.48550/ARXIV.2304.02014}.
\newblock URL \url{https://doi.org/10.48550/arXiv.2304.02014}.

\bibitem[Du et~al.(2024)Du, Luu, Ji, and Ng]{du2024mercury}
Du, M., Luu, A.~T., Ji, B., and Ng, S.-K.
\newblock Mercury: An efficiency benchmark for llm code synthesis.
\newblock \emph{arXiv preprint arXiv:2402.07844}, 2024.

\bibitem[Fan et~al.(2023)Fan, Gokkaya, Harman, Lyubarskiy, Sengupta, Yoo, and Zhang]{fan2023large}
Fan, A., Gokkaya, B., Harman, M., Lyubarskiy, M., Sengupta, S., Yoo, S., and Zhang, J.~M.
\newblock Large language models for software engineering: Survey and open problems.
\newblock In \emph{2023 IEEE/ACM International Conference on Software Engineering: Future of Software Engineering (ICSE-FoSE)}, pp.\  31--53. IEEE, 2023.

\bibitem[Fried et~al.(2023)Fried, Aghajanyan, Lin, Wang, Wallace, Shi, Zhong, Yih, Zettlemoyer, and Lewis]{FriedAL0WSZYZL23}
Fried, D., Aghajanyan, A., Lin, J., Wang, S., Wallace, E., Shi, F., Zhong, R., Yih, S., Zettlemoyer, L., and Lewis, M.
\newblock Incoder: {A} generative model for code infilling and synthesis.
\newblock In \emph{The Eleventh International Conference on Learning Representations, {ICLR} 2023, Kigali, Rwanda, May 1-5, 2023}. OpenReview.net, 2023.
\newblock URL \url{https://openreview.net/pdf?id=hQwb-lbM6EL}.

\bibitem[Gao et~al.(2023)Gao, Dai, Pasupat, Chen, Chaganty, Fan, Zhao, Lao, Lee, Juan, and Guu]{GaoRARR23}
Gao, L., Dai, Z., Pasupat, P., Chen, A., Chaganty, A.~T., Fan, Y., Zhao, V.~Y., Lao, N., Lee, H., Juan, D., and Guu, K.
\newblock {RARR:} researching and revising what language models say, using language models.
\newblock In Rogers, A., Boyd{-}Graber, J.~L., and Okazaki, N. (eds.), \emph{Proceedings of the 61st Annual Meeting of the Association for Computational Linguistics (Volume 1: Long Papers), {ACL} 2023, Toronto, Canada, July 9-14, 2023}, pp.\  16477--16508. Association for Computational Linguistics, 2023.
\newblock \doi{10.18653/V1/2023.ACL-LONG.910}.
\newblock URL \url{https://doi.org/10.18653/v1/2023.acl-long.910}.

\bibitem[Glaese et~al.(2022)Glaese, McAleese, Trebacz, Aslanides, Firoiu, Ewalds, Rauh, Weidinger, Chadwick, Thacker, Campbell{-}Gillingham, Uesato, Huang, Comanescu, Yang, See, Dathathri, Greig, Chen, Fritz, Elias, Green, Mokr{\'{a}}, Fernando, Wu, Foley, Young, Gabriel, Isaac, Mellor, Hassabis, Kavukcuoglu, Hendricks, and Irving]{Glaese2022}
Glaese, A., McAleese, N., Trebacz, M., Aslanides, J., Firoiu, V., Ewalds, T., Rauh, M., Weidinger, L., Chadwick, M.~J., Thacker, P., Campbell{-}Gillingham, L., Uesato, J., Huang, P., Comanescu, R., Yang, F., See, A., Dathathri, S., Greig, R., Chen, C., Fritz, D., Elias, J.~S., Green, R., Mokr{\'{a}}, S., Fernando, N., Wu, B., Foley, R., Young, S., Gabriel, I., Isaac, W., Mellor, J., Hassabis, D., Kavukcuoglu, K., Hendricks, L.~A., and Irving, G.
\newblock Improving alignment of dialogue agents via targeted human judgements.
\newblock \emph{CoRR}, abs/2209.14375, 2022.
\newblock \doi{10.48550/ARXIV.2209.14375}.
\newblock URL \url{https://doi.org/10.48550/arXiv.2209.14375}.

\bibitem[Gou et~al.(2023)Gou, Shao, Gong, Shen, Yang, Duan, and Chen]{GouCritic2023}
Gou, Z., Shao, Z., Gong, Y., Shen, Y., Yang, Y., Duan, N., and Chen, W.
\newblock {CRITIC:} large language models can self-correct with tool-interactive critiquing.
\newblock \emph{CoRR}, abs/2305.11738, 2023.
\newblock \doi{10.48550/ARXIV.2305.11738}.
\newblock URL \url{https://doi.org/10.48550/arXiv.2305.11738}.

\bibitem[Gunasekar et~al.(2023)Gunasekar, Zhang, Aneja, Mendes, Giorno, Gopi, Javaheripi, Kauffmann, de~Rosa, Saarikivi, Salim, Shah, Behl, Wang, Bubeck, Eldan, Kalai, Lee, and Li]{Gunasekar2023}
Gunasekar, S., Zhang, Y., Aneja, J., Mendes, C. C.~T., Giorno, A.~D., Gopi, S., Javaheripi, M., Kauffmann, P., de~Rosa, G., Saarikivi, O., Salim, A., Shah, S., Behl, H.~S., Wang, X., Bubeck, S., Eldan, R., Kalai, A.~T., Lee, Y.~T., and Li, Y.
\newblock Textbooks are all you need.
\newblock \emph{CoRR}, abs/2306.11644, 2023.
\newblock \doi{10.48550/ARXIV.2306.11644}.
\newblock URL \url{https://doi.org/10.48550/arXiv.2306.11644}.

\bibitem[Guo et~al.(2024)Guo, Zhu, Yang, Xie, Dong, Zhang, Chen, Bi, Wu, Li, et~al.]{guo2024deepseek}
Guo, D., Zhu, Q., Yang, D., Xie, Z., Dong, K., Zhang, W., Chen, G., Bi, X., Wu, Y., Li, Y., et~al.
\newblock Deepseek-coder: When the large language model meets programming--the rise of code intelligence.
\newblock \emph{arXiv preprint arXiv:2401.14196}, 2024.

\bibitem[Haque et~al.(2022)Haque, Ahmad, Lourentzou, and Brown]{Haque2022}
Haque, M. M.~A., Ahmad, W.~U., Lourentzou, I., and Brown, C.
\newblock Fixeval: Execution-based evaluation of program fixes for competitive programming problems.
\newblock \emph{CoRR}, abs/2206.07796, 2022.
\newblock \doi{10.48550/ARXIV.2206.07796}.
\newblock URL \url{https://doi.org/10.48550/arXiv.2206.07796}.

\bibitem[Hasan et~al.(2021)Hasan, Muttaqueen, Ishtiaq, Mehrab, Haque, Hasan, Ahmad, Iqbal, and Shahriyar]{HasanMIMHHAIS21}
Hasan, M., Muttaqueen, T., Ishtiaq, A.~A., Mehrab, K.~S., Haque, M. M.~A., Hasan, T., Ahmad, W.~U., Iqbal, A., and Shahriyar, R.
\newblock Codesc: {A} large code-description parallel dataset.
\newblock In Zong, C., Xia, F., Li, W., and Navigli, R. (eds.), \emph{Findings of the Association for Computational Linguistics: {ACL/IJCNLP} 2021, Online Event, August 1-6, 2021}, volume {ACL/IJCNLP} 2021 of \emph{Findings of {ACL}}, pp.\  210--218. Association for Computational Linguistics, 2021.
\newblock \doi{10.18653/V1/2021.FINDINGS-ACL.18}.
\newblock URL \url{https://doi.org/10.18653/v1/2021.findings-acl.18}.

\bibitem[Huang et~al.(2024)Huang, Zhang, Qing, and Cui]{huang2024effibench}
Huang, D., Zhang, J.~M., Qing, Y., and Cui, H.
\newblock Effibench: Benchmarking the efficiency of automatically generated code.
\newblock \emph{arXiv preprint arXiv:2402.02037}, 2024.

\bibitem[Jiang et~al.(2023{\natexlab{a}})Jiang, Liu, Lutellier, and Tan]{JiangLLT23}
Jiang, N., Liu, K., Lutellier, T., and Tan, L.
\newblock Impact of code language models on automated program repair.
\newblock In \emph{45th {IEEE/ACM} International Conference on Software Engineering, {ICSE} 2023, Melbourne, Australia, May 14-20, 2023}, pp.\  1430--1442. {IEEE}, 2023{\natexlab{a}}.
\newblock \doi{10.1109/ICSE48619.2023.00125}.
\newblock URL \url{https://doi.org/10.1109/ICSE48619.2023.00125}.

\bibitem[Jiang et~al.(2023{\natexlab{b}})Jiang, Wang, and Wang]{JiangSelfEvolve23}
Jiang, S., Wang, Y., and Wang, Y.
\newblock Selfevolve: {A} code evolution framework via large language models.
\newblock \emph{CoRR}, abs/2306.02907, 2023{\natexlab{b}}.
\newblock \doi{10.48550/ARXIV.2306.02907}.
\newblock URL \url{https://doi.org/10.48550/arXiv.2306.02907}.

\bibitem[Kreutzer et~al.(2018)Kreutzer, Khadivi, Matusov, and Riezler]{KreutzerKMR18}
Kreutzer, J., Khadivi, S., Matusov, E., and Riezler, S.
\newblock Can neural machine translation be improved with user feedback?
\newblock In Bangalore, S., Chu{-}Carroll, J., and Li, Y. (eds.), \emph{Proceedings of the 2018 Conference of the North American Chapter of the Association for Computational Linguistics: Human Language Technologies, {NAACL-HLT} 2018, New Orleans, Louisiana, USA, June 1-6, 2018, Volume 3 (Industry Papers)}, pp.\  92--105. Association for Computational Linguistics, 2018.
\newblock \doi{10.18653/V1/N18-3012}.
\newblock URL \url{https://doi.org/10.18653/v1/n18-3012}.

\bibitem[Lemieux et~al.(2023)Lemieux, Inala, Lahiri, and Sen]{LemieuxILS23}
Lemieux, C., Inala, J.~P., Lahiri, S.~K., and Sen, S.
\newblock Codamosa: Escaping coverage plateaus in test generation with pre-trained large language models.
\newblock In \emph{45th {IEEE/ACM} International Conference on Software Engineering, {ICSE} 2023, Melbourne, Australia, May 14-20, 2023}, pp.\  919--931. {IEEE}, 2023.
\newblock \doi{10.1109/ICSE48619.2023.00085}.
\newblock URL \url{https://doi.org/10.1109/ICSE48619.2023.00085}.

\bibitem[Li et~al.(2023{\natexlab{a}})Li, Allal, Zi, Muennighoff, Kocetkov, Mou, Marone, Akiki, Li, Chim, Liu, Zheltonozhskii, Zhuo, Wang, Dehaene, Davaadorj, Lamy{-}Poirier, Monteiro, Shliazhko, Gontier, Meade, Zebaze, Yee, Umapathi, Zhu, Lipkin, Oblokulov, Wang, V, Stillerman, Patel, Abulkhanov, Zocca, Dey, Zhang, Moustafa{-}Fahmy, Bhattacharyya, Yu, Singh, Luccioni, Villegas, Kunakov, Zhdanov, Romero, Lee, Timor, Ding, Schlesinger, Schoelkopf, Ebert, Dao, Mishra, Gu, Robinson, Anderson, Dolan{-}Gavitt, Contractor, Reddy, Fried, Bahdanau, Jernite, Ferrandis, Hughes, Wolf, Guha, von Werra, and de~Vries]{LiStarCoder203}
Li, R., Allal, L.~B., Zi, Y., Muennighoff, N., Kocetkov, D., Mou, C., Marone, M., Akiki, C., Li, J., Chim, J., Liu, Q., Zheltonozhskii, E., Zhuo, T.~Y., Wang, T., Dehaene, O., Davaadorj, M., Lamy{-}Poirier, J., Monteiro, J., Shliazhko, O., Gontier, N., Meade, N., Zebaze, A., Yee, M., Umapathi, L.~K., Zhu, J., Lipkin, B., Oblokulov, M., Wang, Z., V, R.~M., Stillerman, J., Patel, S.~S., Abulkhanov, D., Zocca, M., Dey, M., Zhang, Z., Moustafa{-}Fahmy, N., Bhattacharyya, U., Yu, W., Singh, S., Luccioni, S., Villegas, P., Kunakov, M., Zhdanov, F., Romero, M., Lee, T., Timor, N., Ding, J., Schlesinger, C., Schoelkopf, H., Ebert, J., Dao, T., Mishra, M., Gu, A., Robinson, J., Anderson, C.~J., Dolan{-}Gavitt, B., Contractor, D., Reddy, S., Fried, D., Bahdanau, D., Jernite, Y., Ferrandis, C.~M., Hughes, S., Wolf, T., Guha, A., von Werra, L., and de~Vries, H.
\newblock Starcoder: may the source be with you!
\newblock \emph{CoRR}, abs/2305.06161, 2023{\natexlab{a}}.
\newblock \doi{10.48550/ARXIV.2305.06161}.
\newblock URL \url{https://doi.org/10.48550/arXiv.2305.06161}.

\bibitem[Li et~al.(2022)Li, Choi, Chung, Kushman, Schrittwieser, Leblond, Eccles, Keeling, Gimeno, Lago, Hubert, Choy, de~Masson~d'Autume, Babuschkin, Chen, Huang, Welbl, Gowal, Cherepanov, Molloy, Mankowitz, Robson, Kohli, de~Freitas, Kavukcuoglu, and Vinyals]{Lialphacode2022}
Li, Y., Choi, D.~H., Chung, J., Kushman, N., Schrittwieser, J., Leblond, R., Eccles, T., Keeling, J., Gimeno, F., Lago, A.~D., Hubert, T., Choy, P., de~Masson~d'Autume, C., Babuschkin, I., Chen, X., Huang, P., Welbl, J., Gowal, S., Cherepanov, A., Molloy, J., Mankowitz, D.~J., Robson, E.~S., Kohli, P., de~Freitas, N., Kavukcuoglu, K., and Vinyals, O.
\newblock Competition-level code generation with alphacode.
\newblock \emph{CoRR}, abs/2203.07814, 2022.
\newblock \doi{10.48550/ARXIV.2203.07814}.
\newblock URL \url{https://doi.org/10.48550/arXiv.2203.07814}.

\bibitem[Li et~al.(2023{\natexlab{b}})Li, Bubeck, Eldan, Giorno, Gunasekar, and Lee]{Liphi2023}
Li, Y., Bubeck, S., Eldan, R., Giorno, A.~D., Gunasekar, S., and Lee, Y.~T.
\newblock Textbooks are all you need {II:} phi-1.5 technical report.
\newblock \emph{CoRR}, abs/2309.05463, 2023{\natexlab{b}}.
\newblock \doi{10.48550/ARXIV.2309.05463}.
\newblock URL \url{https://doi.org/10.48550/arXiv.2309.05463}.

\bibitem[Liu et~al.(2024{\natexlab{a}})Liu, Xia, Wang, and Zhang]{liu2024your}
Liu, J., Xia, C.~S., Wang, Y., and Zhang, L.
\newblock Is your code generated by chatgpt really correct? rigorous evaluation of large language models for code generation.
\newblock \emph{Advances in Neural Information Processing Systems}, 36, 2024{\natexlab{a}}.

\bibitem[Liu et~al.(2024{\natexlab{b}})Liu, Xie, Wang, Wei, Ding, and Zhang]{liu2024evaluating}
Liu, J., Xie, S., Wang, J., Wei, Y., Ding, Y., and Zhang, L.
\newblock Evaluating language models for efficient code generation.
\newblock \emph{arXiv preprint arXiv:2408.06450}, 2024{\natexlab{b}}.

\bibitem[Lu et~al.(2024{\natexlab{a}})Lu, Dou, Wang, Cao, Dai, Wan, Huang, and Guo]{Lu2024Auto}
Lu, J., Dou, Z., Wang, H., Cao, Z., Dai, J., Wan, Y., Huang, Y., and Guo, Z.
\newblock Autocv: Empowering reasoning with automated process labeling via confidence variation.
\newblock \emph{CoRR}, abs/2405.16802, 2024{\natexlab{a}}.
\newblock \doi{10.48550/ARXIV.2405.16802}.
\newblock URL \url{https://doi.org/10.48550/arXiv.2405.16802}.

\bibitem[Lu et~al.(2024{\natexlab{b}})Lu, Liu, Wan, Huang, Wang, Yang, Tang, and Guo]{Lu2024pda}
Lu, J., Liu, Z., Wan, Y., Huang, Y., Wang, H., Yang, Z., Tang, J., and Guo, Z.
\newblock Process-driven autoformalization in lean 4.
\newblock \emph{CoRR}, abs/2406.01940, 2024{\natexlab{b}}.
\newblock \doi{10.48550/ARXIV.2406.01940}.
\newblock URL \url{https://doi.org/10.48550/arXiv.2406.01940}.

\bibitem[Luo et~al.(2023)Luo, Xu, Zhao, Sun, Geng, Hu, Tao, Ma, Lin, and Jiang]{luo2023wizardcoder}
Luo, Z., Xu, C., Zhao, P., Sun, Q., Geng, X., Hu, W., Tao, C., Ma, J., Lin, Q., and Jiang, D.
\newblock Wizardcoder: Empowering code large language models with evol-instruct.
\newblock \emph{arXiv preprint arXiv:2306.08568}, 2023.

\bibitem[Luo et~al.(2024)Luo, Xu, Zhao, Sun, Geng, Hu, Tao, Ma, Lin, and Jiang]{WizardCoder2024}
Luo, Z., Xu, C., Zhao, P., Sun, Q., Geng, X., Hu, W., Tao, C., Ma, J., Lin, Q., and Jiang, D.
\newblock Wizardcoder: Empowering code large language models with evol-instruct.
\newblock In \emph{The Twelfth International Conference on Learning Representations, {ICLR} 2024, Vienna, Austria, May 7-11, 2024}. OpenReview.net, 2024.
\newblock URL \url{https://openreview.net/forum?id=UnUwSIgK5W}.

\bibitem[Madaan et~al.(2023)Madaan, Tandon, Gupta, Hallinan, Gao, Wiegreffe, Alon, Dziri, Prabhumoye, Yang, Gupta, Majumder, Hermann, Welleck, Yazdanbakhsh, and Clark]{MadaanSelfRefine23}
Madaan, A., Tandon, N., Gupta, P., Hallinan, S., Gao, L., Wiegreffe, S., Alon, U., Dziri, N., Prabhumoye, S., Yang, Y., Gupta, S., Majumder, B.~P., Hermann, K., Welleck, S., Yazdanbakhsh, A., and Clark, P.
\newblock Self-refine: Iterative refinement with self-feedback.
\newblock In Oh, A., Naumann, T., Globerson, A., Saenko, K., Hardt, M., and Levine, S. (eds.), \emph{Advances in Neural Information Processing Systems 36: Annual Conference on Neural Information Processing Systems 2023, NeurIPS 2023, New Orleans, LA, USA, December 10 - 16, 2023}, 2023.
\newblock URL \url{http://papers.nips.cc/paper\_files/paper/2023/hash/91edff07232fb1b55a505a9e9f6c0ff3-Abstract-Conference.html}.

\bibitem[Mancebo et~al.(2021)Mancebo, Garcia, and Calero]{mancebo2021process}
Mancebo, J., Garcia, F., and Calero, C.
\newblock A process for analysing the energy efficiency of software.
\newblock \emph{Information and Software Technology}, 134:\penalty0 106560, 2021.

\bibitem[Meta(2024)]{Llama3}
Meta.
\newblock Introducing meta llama 3: The most capable openly available llm to date, 2024.
\newblock URL \url{https://ai.meta.com/blog/meta-llama-3/}.

\bibitem[Metcalfe(2017)]{metcalfe2017learning}
Metcalfe, J.
\newblock Learning from errors.
\newblock \emph{Annual review of psychology}, 68:\penalty0 465--489, 2017.

\bibitem[Microsoft(2024)]{Copilot}
Microsoft.
\newblock The world’s most widely adopted ai developer tool., 2024.
\newblock URL \url{https://github.com/features/copilot}.

\bibitem[Mir et~al.(2022)Mir, Latoskinas, Proksch, and Gousios]{MirLPG22}
Mir, A.~M., Latoskinas, E., Proksch, S., and Gousios, G.
\newblock Type4py: Practical deep similarity learning-based type inference for python.
\newblock In \emph{44th {IEEE/ACM} 44th International Conference on Software Engineering, {ICSE} 2022, Pittsburgh, PA, USA, May 25-27, 2022}, pp.\  2241--2252. {ACM}, 2022.
\newblock \doi{10.1145/3510003.3510124}.
\newblock URL \url{https://doi.org/10.1145/3510003.3510124}.

\bibitem[Muennighoff et~al.(2024)Muennighoff, Liu, Zebaze, Zheng, Hui, Zhuo, Singh, Tang, von Werra, and Longpre]{OctoPack2024}
Muennighoff, N., Liu, Q., Zebaze, A.~R., Zheng, Q., Hui, B., Zhuo, T.~Y., Singh, S., Tang, X., von Werra, L., and Longpre, S.
\newblock Octopack: Instruction tuning code large language models.
\newblock In \emph{The Twelfth International Conference on Learning Representations, {ICLR} 2024, Vienna, Austria, May 7-11, 2024}. OpenReview.net, 2024.
\newblock URL \url{https://openreview.net/forum?id=mw1PWNSWZP}.

\bibitem[Nijkamp et~al.(2023)Nijkamp, Pang, Hayashi, Tu, Wang, Zhou, Savarese, and Xiong]{NijkampPHTWZSX23}
Nijkamp, E., Pang, B., Hayashi, H., Tu, L., Wang, H., Zhou, Y., Savarese, S., and Xiong, C.
\newblock Codegen: An open large language model for code with multi-turn program synthesis.
\newblock In \emph{The Eleventh International Conference on Learning Representations, {ICLR} 2023, Kigali, Rwanda, May 1-5, 2023}. OpenReview.net, 2023.
\newblock URL \url{https://openreview.net/pdf?id=iaYcJKpY2B\_}.

\bibitem[Niu et~al.(2024)Niu, Zhang, Li, Luo, and Ng]{niu2024evaluating}
Niu, C., Zhang, T., Li, C., Luo, B., and Ng, V.
\newblock On evaluating the efficiency of source code generated by llms.
\newblock \emph{arXiv preprint arXiv:2404.06041}, 2024.

\bibitem[OpenAI(2023{\natexlab{a}})]{GPT35turbo}
OpenAI.
\newblock {GPT-3.5} {T}urbo, 2023{\natexlab{a}}.
\newblock URL \url{https://platform.openai.com/docs/models/gpt-3-5}.

\bibitem[OpenAI(2023{\natexlab{b}})]{GPT4}
OpenAI.
\newblock {GPT-4} {T}echnical {R}eport.
\newblock \emph{CoRR}, abs/2303.08774, 2023{\natexlab{b}}.
\newblock \doi{10.48550/arXiv.2303.08774}.
\newblock URL \url{https://doi.org/10.48550/arXiv.2303.08774}.

\bibitem[Ouyang et~al.(2022)Ouyang, Wu, Jiang, Almeida, Wainwright, Mishkin, Zhang, Agarwal, Slama, Ray, Schulman, Hilton, Kelton, Miller, Simens, Askell, Welinder, Christiano, Leike, and Lowe]{Ouyang0JAWMZASR22}
Ouyang, L., Wu, J., Jiang, X., Almeida, D., Wainwright, C.~L., Mishkin, P., Zhang, C., Agarwal, S., Slama, K., Ray, A., Schulman, J., Hilton, J., Kelton, F., Miller, L., Simens, M., Askell, A., Welinder, P., Christiano, P.~F., Leike, J., and Lowe, R.
\newblock Training language models to follow instructions with human feedback.
\newblock In Koyejo, S., Mohamed, S., Agarwal, A., Belgrave, D., Cho, K., and Oh, A. (eds.), \emph{Advances in Neural Information Processing Systems 35: Annual Conference on Neural Information Processing Systems 2022, NeurIPS 2022, New Orleans, LA, USA, November 28 - December 9, 2022}, 2022.
\newblock URL \url{http://papers.nips.cc/paper\_files/paper/2022/hash/b1efde53be364a73914f58805a001731-Abstract-Conference.html}.

\bibitem[Pereira et~al.(2021)Pereira, Couto, Ribeiro, Rua, Cunha, Fernandes, and Saraiva]{pereira2021ranking}
Pereira, R., Couto, M., Ribeiro, F., Rua, R., Cunha, J., Fernandes, J.~P., and Saraiva, J.
\newblock Ranking programming languages by energy efficiency.
\newblock \emph{Science of Computer Programming}, 205:\penalty0 102609, 2021.

\bibitem[Qiu et~al.(2024)Qiu, Zeng, Tong, Ezick, and Lott]{qiu2024efficient}
Qiu, R., Zeng, W.~W., Tong, H., Ezick, J., and Lott, C.
\newblock How efficient is llm-generated code? a rigorous \& high-standard benchmark.
\newblock \emph{arXiv preprint arXiv:2406.06647}, 2024.

\bibitem[Rozi{\`{e}}re et~al.(2020)Rozi{\`{e}}re, Lachaux, Chanussot, and Lample]{RoziereLCL20}
Rozi{\`{e}}re, B., Lachaux, M., Chanussot, L., and Lample, G.
\newblock Unsupervised translation of programming languages.
\newblock In Larochelle, H., Ranzato, M., Hadsell, R., Balcan, M., and Lin, H. (eds.), \emph{Advances in Neural Information Processing Systems 33: Annual Conference on Neural Information Processing Systems 2020, NeurIPS 2020, December 6-12, 2020, virtual}, 2020.
\newblock URL \url{https://proceedings.neurips.cc/paper/2020/hash/ed23fbf18c2cd35f8c7f8de44f85c08d-Abstract.html}.

\bibitem[Rozi{\`{e}}re et~al.(2023)Rozi{\`{e}}re, Gehring, Gloeckle, Sootla, Gat, Tan, Adi, Liu, Remez, Rapin, Kozhevnikov, Evtimov, Bitton, Bhatt, Canton{-}Ferrer, Grattafiori, Xiong, D{\'{e}}fossez, Copet, Azhar, Touvron, Martin, Usunier, Scialom, and Synnaeve]{Roziere2023}
Rozi{\`{e}}re, B., Gehring, J., Gloeckle, F., Sootla, S., Gat, I., Tan, X.~E., Adi, Y., Liu, J., Remez, T., Rapin, J., Kozhevnikov, A., Evtimov, I., Bitton, J., Bhatt, M., Canton{-}Ferrer, C., Grattafiori, A., Xiong, W., D{\'{e}}fossez, A., Copet, J., Azhar, F., Touvron, H., Martin, L., Usunier, N., Scialom, T., and Synnaeve, G.
\newblock Code llama: Open foundation models for code.
\newblock \emph{CoRR}, abs/2308.12950, 2023.
\newblock \doi{10.48550/ARXIV.2308.12950}.
\newblock URL \url{https://doi.org/10.48550/arXiv.2308.12950}.

\bibitem[Shi et~al.(2024)Shi, Yang, and Lo]{shi2024efficient}
Shi, J., Yang, Z., and Lo, D.
\newblock Efficient and green large language models for software engineering: Vision and the road ahead.
\newblock \emph{arXiv preprint arXiv:2404.04566}, 2024.

\bibitem[Shinn et~al.(2023)Shinn, Cassano, Gopinath, Narasimhan, and Yao]{ShinnReflexion23}
Shinn, N., Cassano, F., Gopinath, A., Narasimhan, K., and Yao, S.
\newblock Reflexion: language agents with verbal reinforcement learning.
\newblock In Oh, A., Naumann, T., Globerson, A., Saenko, K., Hardt, M., and Levine, S. (eds.), \emph{Advances in Neural Information Processing Systems 36: Annual Conference on Neural Information Processing Systems 2023, NeurIPS 2023, New Orleans, LA, USA, December 10 - 16, 2023}, 2023.
\newblock URL \url{http://papers.nips.cc/paper\_files/paper/2023/hash/1b44b878bb782e6954cd888628510e90-Abstract-Conference.html}.

\bibitem[Shypula et~al.(2024)Shypula, Madaan, Zeng, Alon, Gardner, Hashemi, Neubig, Ranganathan, Bastani, and Yazdanbakhsh]{pie_iclr_2024_spotlight}
Shypula, A., Madaan, A., Zeng, Y., Alon, U., Gardner, J., Hashemi, M., Neubig, G., Ranganathan, P., Bastani, O., and Yazdanbakhsh, A.
\newblock \href{https://openreview.net/pdf?id=ix7rLVHXyY}{Learning Performance-Improving Code Edits}.
\newblock In \emph{The Twelfth International Conference on Learning Representations (ICLR)}, 2024.

\bibitem[Team(2023)]{xwin-lm}
Team, X.-L.
\newblock Xwin-lm, 9 2023.
\newblock URL \url{https://github.com/Xwin-LM/Xwin-LM}.

\bibitem[Touvron et~al.(2023)Touvron, Martin, Stone, Albert, Almahairi, Babaei, Bashlykov, Batra, Bhargava, Bhosale, Bikel, Blecher, Canton{-}Ferrer, Chen, Cucurull, Esiobu, Fernandes, Fu, Fu, Fuller, Gao, Goswami, Goyal, Hartshorn, Hosseini, Hou, Inan, Kardas, Kerkez, Khabsa, Kloumann, Korenev, Koura, Lachaux, Lavril, Lee, Liskovich, Lu, Mao, Martinet, Mihaylov, Mishra, Molybog, Nie, Poulton, Reizenstein, Rungta, Saladi, Schelten, Silva, Smith, Subramanian, Tan, Tang, Taylor, Williams, Kuan, Xu, Yan, Zarov, Zhang, Fan, Kambadur, Narang, Rodriguez, Stojnic, Edunov, and Scialom]{Touvron2023}
Touvron, H., Martin, L., Stone, K., Albert, P., Almahairi, A., Babaei, Y., Bashlykov, N., Batra, S., Bhargava, P., Bhosale, S., Bikel, D., Blecher, L., Canton{-}Ferrer, C., Chen, M., Cucurull, G., Esiobu, D., Fernandes, J., Fu, J., Fu, W., Fuller, B., Gao, C., Goswami, V., Goyal, N., Hartshorn, A., Hosseini, S., Hou, R., Inan, H., Kardas, M., Kerkez, V., Khabsa, M., Kloumann, I., Korenev, A., Koura, P.~S., Lachaux, M., Lavril, T., Lee, J., Liskovich, D., Lu, Y., Mao, Y., Martinet, X., Mihaylov, T., Mishra, P., Molybog, I., Nie, Y., Poulton, A., Reizenstein, J., Rungta, R., Saladi, K., Schelten, A., Silva, R., Smith, E.~M., Subramanian, R., Tan, X.~E., Tang, B., Taylor, R., Williams, A., Kuan, J.~X., Xu, P., Yan, Z., Zarov, I., Zhang, Y., Fan, A., Kambadur, M., Narang, S., Rodriguez, A., Stojnic, R., Edunov, S., and Scialom, T.
\newblock Llama 2: Open foundation and fine-tuned chat models.
\newblock \emph{CoRR}, abs/2307.09288, 2023.
\newblock \doi{10.48550/ARXIV.2307.09288}.
\newblock URL \url{https://doi.org/10.48550/arXiv.2307.09288}.

\bibitem[Viswanathan et~al.(2023)Viswanathan, Zhao, Bertsch, Wu, and Neubig]{viswanathan2023prompt2model}
Viswanathan, V., Zhao, C., Bertsch, A., Wu, T., and Neubig, G.
\newblock Prompt2model: Generating deployable models from natural language instructions.
\newblock \emph{arXiv preprint arXiv:2308.12261}, 2023.

\bibitem[Waghjale et~al.(2024)Waghjale, Veerendranath, Wang, and Fried]{waghjale2024ecco}
Waghjale, S., Veerendranath, V., Wang, Z.~Z., and Fried, D.
\newblock Ecco: Can we improve model-generated code efficiency without sacrificing functional correctness?
\newblock \emph{arXiv preprint arXiv:2407.14044}, 2024.

\bibitem[Wang et~al.(2021)Wang, Wang, Joty, and Hoi]{WangCodeT52021}
Wang, Y., Wang, W., Joty, S.~R., and Hoi, S. C.~H.
\newblock Codet5: Identifier-aware unified pre-trained encoder-decoder models for code understanding and generation.
\newblock In Moens, M., Huang, X., Specia, L., and Yih, S.~W. (eds.), \emph{Proceedings of the 2021 Conference on Empirical Methods in Natural Language Processing, {EMNLP} 2021, Virtual Event / Punta Cana, Dominican Republic, 7-11 November, 2021}, pp.\  8696--8708. Association for Computational Linguistics, 2021.
\newblock \doi{10.18653/V1/2021.EMNLP-MAIN.685}.
\newblock URL \url{https://doi.org/10.18653/v1/2021.emnlp-main.685}.

\bibitem[Wang et~al.(2023)Wang, Kordi, Mishra, Liu, Smith, Khashabi, and Hajishirzi]{WangKMLSKH23}
Wang, Y., Kordi, Y., Mishra, S., Liu, A., Smith, N.~A., Khashabi, D., and Hajishirzi, H.
\newblock Self-instruct: Aligning language models with self-generated instructions.
\newblock In Rogers, A., Boyd{-}Graber, J.~L., and Okazaki, N. (eds.), \emph{Proceedings of the 61st Annual Meeting of the Association for Computational Linguistics (Volume 1: Long Papers), {ACL} 2023, Toronto, Canada, July 9-14, 2023}, pp.\  13484--13508. Association for Computational Linguistics, 2023.
\newblock \doi{10.18653/V1/2023.ACL-LONG.754}.
\newblock URL \url{https://doi.org/10.18653/v1/2023.acl-long.754}.

\bibitem[Wei et~al.(2022)Wei, Bosma, Zhao, Guu, Yu, Lester, Du, Dai, and Le]{WeiBZGYLDDL22}
Wei, J., Bosma, M., Zhao, V.~Y., Guu, K., Yu, A.~W., Lester, B., Du, N., Dai, A.~M., and Le, Q.~V.
\newblock Finetuned language models are zero-shot learners.
\newblock In \emph{The Tenth International Conference on Learning Representations, {ICLR} 2022, Virtual Event, April 25-29, 2022}. OpenReview.net, 2022.
\newblock URL \url{https://openreview.net/forum?id=gEZrGCozdqR}.

\bibitem[Wei et~al.(2023)Wei, Durrett, and Dillig]{WeiDD23}
Wei, J., Durrett, G., and Dillig, I.
\newblock Typet5: Seq2seq type inference using static analysis.
\newblock In \emph{The Eleventh International Conference on Learning Representations, {ICLR} 2023, Kigali, Rwanda, May 1-5, 2023}. OpenReview.net, 2023.
\newblock URL \url{https://openreview.net/pdf?id=4TyNEhI2GdN}.

\bibitem[Wei et~al.(2024)Wei, Wang, Liu, Ding, and Zhang]{Magicoder2024}
Wei, Y., Wang, Z., Liu, J., Ding, Y., and Zhang, L.
\newblock Magicoder: Empowering code generation with oss-instruct.
\newblock In \emph{Forty-first International Conference on Machine Learning, {ICML} 2024, Vienna, Austria, July 21-27, 2024}. OpenReview.net, 2024.
\newblock URL \url{https://openreview.net/forum?id=XUeoOBid3x}.

\bibitem[Xu et~al.(2024)Xu, Sun, Zheng, Geng, Zhao, Feng, Tao, Lin, and Jiang]{WizardLM2024}
Xu, C., Sun, Q., Zheng, K., Geng, X., Zhao, P., Feng, J., Tao, C., Lin, Q., and Jiang, D.
\newblock Wizardlm: Empowering large pre-trained language models to follow complex instructions.
\newblock In \emph{The Twelfth International Conference on Learning Representations, {ICLR} 2024, Vienna, Austria, May 7-11, 2024}. OpenReview.net, 2024.
\newblock URL \url{https://openreview.net/forum?id=CfXh93NDgH}.

\bibitem[Yao et~al.(2024)Yao, Wu, Guo, Zhou, Gao, Luo, Hou, Fu, and Song]{Yao2024learning}
Yao, Y., Wu, H., Guo, Z., Zhou, B., Gao, J., Luo, S., Hou, H., Fu, X., and Song, L.
\newblock Learning from correctness without prompting makes {LLM} efficient reasoner.
\newblock \emph{CoRR}, abs/2403.19094, 2024.
\newblock \doi{10.48550/ARXIV.2403.19094}.
\newblock URL \url{https://doi.org/10.48550/arXiv.2403.19094}.

\bibitem[Yu et~al.(2023)Yu, Zhang, Liang, Jiang, and Sabharwal]{YuReFeed23}
Yu, W., Zhang, Z., Liang, Z., Jiang, M., and Sabharwal, A.
\newblock Improving language models via plug-and-play retrieval feedback.
\newblock \emph{CoRR}, abs/2305.14002, 2023.
\newblock \doi{10.48550/ARXIV.2305.14002}.
\newblock URL \url{https://doi.org/10.48550/arXiv.2305.14002}.

\bibitem[Zhang et~al.(2023{\natexlab{a}})Zhang, Li, Li, Li, and Jin]{ZhangSelfEdit23}
Zhang, K., Li, Z., Li, J., Li, G., and Jin, Z.
\newblock Self-edit: Fault-aware code editor for code generation.
\newblock In Rogers, A., Boyd{-}Graber, J.~L., and Okazaki, N. (eds.), \emph{Proceedings of the 61st Annual Meeting of the Association for Computational Linguistics (Volume 1: Long Papers), {ACL} 2023, Toronto, Canada, July 9-14, 2023}, pp.\  769--787. Association for Computational Linguistics, 2023{\natexlab{a}}.
\newblock \doi{10.18653/V1/2023.ACL-LONG.45}.
\newblock URL \url{https://doi.org/10.18653/v1/2023.acl-long.45}.

\bibitem[Zhang et~al.(2023{\natexlab{b}})Zhang, Wang, Xia, Wang, and Li]{ZhangAlgo23}
Zhang, K., Wang, D., Xia, J., Wang, W.~Y., and Li, L.
\newblock {ALGO:} synthesizing algorithmic programs with generated oracle verifiers.
\newblock In Oh, A., Naumann, T., Globerson, A., Saenko, K., Hardt, M., and Levine, S. (eds.), \emph{Advances in Neural Information Processing Systems 36: Annual Conference on Neural Information Processing Systems 2023, NeurIPS 2023, New Orleans, LA, USA, December 10 - 16, 2023}, 2023{\natexlab{b}}.
\newblock URL \url{http://papers.nips.cc/paper\_files/paper/2023/hash/abe1eb21ceb046209c96a0f5e7544ccc-Abstract-Conference.html}.

\bibitem[Zhao et~al.(2024)Zhao, Jia, Viswanathan, Wu, and Neubig]{zhao2024self}
Zhao, C., Jia, X., Viswanathan, V., Wu, T., and Neubig, G.
\newblock Self-guide: Better task-specific instruction following via self-synthetic finetuning.
\newblock \emph{arXiv preprint arXiv:2407.12874}, 2024.

\bibitem[Zheng et~al.(2024)Zheng, Zhang, Shen, Liu, Lin, Fu, Chen, and Yue]{zheng2024opencodeinterpreter}
Zheng, T., Zhang, G., Shen, T., Liu, X., Lin, B.~Y., Fu, J., Chen, W., and Yue, X.
\newblock Opencodeinterpreter: Integrating code generation with execution and refinement.
\newblock \emph{arXiv preprint arXiv:2402.14658}, 2024.

\end{thebibliography}
\bibliographystyle{icml2024}

\clearpage
\appendix


\section{Appendix}

\subsection{Limitations}
\label{app:limitations}

\name presents a compelling solution for enhancing the efficiency of code generated by LLMs. However, it's crucial to acknowledge several potential limitations. Firstly, the multi-iteration self-optimization process at the heart of \name can be time-consuming, particularly when applied to intricate programming tasks. Yet, it's important to note that this investment of time can yield significant long-term benefits, as the optimized codes, once deployed, can considerably improve efficiency. Moreover, the process of feeding overhead profiles to LLMs to prompt code optimization may consume more tokens. This is due to the fact that the length of the overhead profiles necessitates additional tokens. Lastly, the effectiveness of EffiLearner has been primarily evaluated on Python. Therefore, its performance in different programming languages or environments may vary, underscoring the need for further testing and validation of this approach in a diverse range of contexts.

\subsection{Broader Impacts}
\label{app:impacts}

\paragraph{Positive Societal Impacts}
The proposed method \name has the potential to bring about several positive societal impacts. Primarily, it can significantly enhance productivity by enabling developers to complete tasks more quickly and efficiently, due to the faster execution times and lower memory and processing power consumption of the optimized code. This is particularly beneficial in resource-constrained environments, such as mobile devices or embedded systems, where conserving resources is crucial. Moreover, \name's focus on improving code efficiency can lead to more cost-effective solutions, as lower resource consumption translates to reduced operational costs.

\paragraph{Negative Societal Impacts}
The increasing effectiveness of LLM-based tools like \name could also have negative societal impacts. For instance, it could lead to an over-reliance on these systems, potentially resulting in a decline in human coding skills and a lack of understanding of the underlying code. This could be problematic in situations where human intervention is necessary. Additionally, there is a risk of job displacement in the coding and programming industry, as the demand for human coders may decrease, particularly for routine coding tasks. Lastly, while \name can improve code efficiency, it does not necessarily address potential security vulnerabilities or privacy issues in the code, which could have significant societal impacts if not properly managed.

\subsection{Prompt Template}
\label{app:template}
\begin{figure*}[htbp]
    \centering
    \begin{minipage}{1\textwidth}

        \begin{minipage}[t]{1\textwidth}
            \begin{modelbox}[title=\textbf{Prompt Template}]
\begin{minted}{python}
prompt = f"""
Please optimize the efficiency of the following Python code based on the task description, test case, and overhead analysis provided. Ensure the optimized code can pass the given test case.

Task Description:
{task_description}

Test Case:
{test_case}

Original Code:
```python
{completion}
```

Overhead Analysis:
{overhead_prompt}
\end{minted}
            \end{modelbox}
        \end{minipage}%
    \end{minipage}

    \caption{Prompt template used by \name in the self-optimization stage.}
    \label{fig:prompt_template}
\end{figure*}

\subsection{Comparison with Baselines}

To demonstrate \name's effectiveness with existing baselines and some prompt engineering methods that can refine the prompt from the correctness to the efficiency of LLM-generated code.

\begin{table}
\tiny
\centering
\caption{Evaluation results of \name and baselines. Since the finetuned link for GPT-3.5-turbo from PIE is not available, we use the fine-tuned CodeLlama 7B for experiments. Due to the fine-tuned PIE CodeLlama 7B does not have the same correct tasks as the original CodeLlama, we then do not provide the initial version for the experiments.}
\begin{tabular}{l|rrrrrr}
\toprule
OptimizationProfile & ET(s) & NET & MU(Mb) & NMU & TMU(Mb*s) & NTMU \\
\midrule
\multicolumn{7}{l}{GPT-3.5-Turbo-0301} \\
\midrule
InitialVersion & 0.36 & 2.50 & 91.25 & 2.45 & 157.50 & 19.75 \\
UnsupervisedSelf-Refine & 0.32 & 2.46 & 78.39 & 2.12 & 312.99 & 42.42 \\
Result-AwareSelf-Refine & 0.30 & 2.25 & 58.65 & 1.61 & 195.49 & 27.16 \\
Self-Edit & 0.42 & 3.67 & 59.86 & 1.65 & 24.87 & 3.28 \\
Critic & 0.60 & 4.25 & 102.75 & 2.82 & 351.91 & 46.39 \\
DirectlyEfficiency & 0.43 & 3.03 & 59.11 & 1.67 & 20.37 & 2.88 \\
Self-RefineEfficiency & 0.40 & 2.83 & 59.11 & 1.67 & 18.80 & 2.65 \\
IsSelf-Refine & 0.40 & 2.88 & 61.83 & 1.81 & 36.29 & 5.69 \\
Self-Reasoning & 0.89 & 6.21 & 60.64 & 1.62 & 45.91 & 5.61 \\
Self-Relfection & 0.81 & 5.67 & 60.64 & 1.62 & 39.35 & 4.80 \\
\name&0.28 \textbf{({\color{problemcolor} 22.2\%})}&2.01 \textbf{({\color{modelcolor} 19.6\%})}&36.08 \textbf{({\color{problemcolor} 60.5\%})}&0.99 \textbf{({\color{modelcolor} 59.6\%})}&12.43 \textbf{({\color{problemcolor} 92.1\%})}&1.64 \textbf{({\color{modelcolor} 91.7\%})}\\
\midrule
\multicolumn{7}{l}{CodeLlama7B(PIE:HQ+SelfPlay)} \\
\midrule
PIE+Zero-Shot & 0.87 & 5.73 & 74.83 & 1.81 & 109.29 & 9.69 \\
PIE+\name+Zero-Shot & 0.79 & 5.41 & 65.78 & 1.68 & 89.90 & 7.84 \\
PIE+Few-Shot & 0.82 & 5.58 & 73.57 & 1.74 & 98.02 & 8.92 \\
PIE+\name+Few-Shot & 0.41 & 2.97 & 73.10 & 1.74 & 59.69 & 5.09 \\
PIE+CoT & 0.79 & 5.14 & 73.14 & 1.74 & 63.93 & 5.35 \\
PIE+\name+COT & 0.45 & 2.84 & 71.15 & 1.71 & 58.06 & 4.77 \\
PIE+DynamicRetrieval,k=4 & 0.74 & 5.36 & 68.64 & 1.51 & 85.24 & 7.78 \\
PIE+\name+DynamicRetrieval,k=4 & 0.41 & 3.36 & 68.63 & 1.51 & 52.34 & 4.52 \\
\midrule
\multicolumn{7}{l}{Supersonic}\\
\midrule
Supersonic & 1.40 & 10.33 & 113.06 & 3.18 & 329.59 & 56.24 \\
Supersonic+\name & 1.34 & 9.91 & 102.26 & 2.87 & 267.47 & 45.64 \\
\bottomrule
\end{tabular}
\label{tab:author_rebuttal_table1}
\end{table}

\begin{table*}[ht!]
\tiny
    \centering
    \caption{Evaluation results of \name's effectiveness in the HumanEval dataset.}
    \begin{tabular}{lrrrrrr}
        \toprule
        Steps&\textbf{ET (s)}&\textbf{NET}&\textbf{MU (Mb)}&\textbf{NMU}&\textbf{TMU (Mb*s)}&\textbf{NTMU}\\
        \midrule
OpenCodeInterpreter-DS-1.3B&0.20&0.86&57.24&1.00&6.63&0.84\\
&0.19 ({\color{problemcolor} 5.0\%})&0.81 ({\color{modelcolor} 5.8\%})&57.17 ({\color{problemcolor} 0.1\%})&1.00 ({\color{modelcolor} 0.0\%})&6.20 ({\color{problemcolor} 6.5\%})&0.79 ({\color{modelcolor} 6.0\%})\\
OpenCodeInterpreter-DS-6.7B&0.21&0.98&58.83&1.06&6.79&0.99\\
&0.21 ({\color{problemcolor} 0.0\%})&0.97 ({\color{modelcolor} 1.0\%})&58.79 ({\color{problemcolor} 0.1\%})&1.06 ({\color{modelcolor} 0.0\%})&6.64 ({\color{problemcolor} 2.2\%})&0.97 ({\color{modelcolor} 2.0\%})\\
OpenCodeInterpreter-DS-33B&0.21&0.95&59.90&1.05&7.05&0.94\\
&0.21 ({\color{problemcolor} 0.0\%})&0.93 ({\color{modelcolor} 2.1\%})&59.93 ({\color{problemcolor} -0.1\%})&1.05 ({\color{modelcolor} 0.0\%})&6.87 ({\color{problemcolor} 2.6\%})&0.92 ({\color{modelcolor} 2.1\%})\\
deepseek-coder-1.3b-instruct&0.23&0.90&62.80&1.00&7.85&0.87\\
&0.22 ({\color{problemcolor} 4.3\%})&0.85 ({\color{modelcolor} 5.6\%})&62.96 ({\color{problemcolor} -0.3\%})&1.00 ({\color{modelcolor} 0.0\%})&7.80 ({\color{problemcolor} 0.6\%})&0.86 ({\color{modelcolor} 1.1\%})\\
deepseek-coder-6.7b-instruct&0.22&0.76&59.57&1.00&7.34&0.77\\
&0.19 ({\color{problemcolor} 13.6\%})&0.68 ({\color{modelcolor} 10.5\%})&59.72 ({\color{problemcolor} -0.3\%})&1.00 ({\color{modelcolor} 0.0\%})&6.59 ({\color{problemcolor} 10.2\%})&0.69 ({\color{modelcolor} 10.4\%})\\
deepseek-coder-33b-instruct&0.21&0.95&63.52&0.99&7.18&0.95\\
&0.20 ({\color{problemcolor} 4.8\%})&0.92 ({\color{modelcolor} 3.2\%})&63.49 ({\color{problemcolor} 0.0\%})&0.99 ({\color{modelcolor} 0.0\%})&6.99 ({\color{problemcolor} 2.6\%})&0.92 ({\color{modelcolor} 3.2\%})\\
CodeLlama-7b-Instruct-hf&0.20&0.71&57.39&0.91&7.08&0.70\\
&0.18 ({\color{problemcolor} 10.0\%})&0.63 ({\color{modelcolor} 11.3\%})&57.07 ({\color{problemcolor} 0.6\%})&0.91 ({\color{modelcolor} 0.0\%})&6.18 ({\color{problemcolor} 12.7\%})&0.61 ({\color{modelcolor} 12.9\%})\\
CodeLlama-13b-Instruct-hf&0.23&0.95&58.13&0.96&7.97&0.94\\
&0.20 ({\color{problemcolor} 13.0\%})&0.80 ({\color{modelcolor} 15.8\%})&58.03 ({\color{problemcolor} 0.2\%})&0.96 ({\color{modelcolor} 0.0\%})&6.64 ({\color{problemcolor} 16.7\%})&0.79 ({\color{modelcolor} 16.0\%})\\
CodeLlama-34b-Instruct-hf&0.24&0.95&61.79&1.01&8.45&0.96\\
&0.21 ({\color{problemcolor} 12.5\%})&0.81 ({\color{modelcolor} 14.7\%})&61.55 ({\color{problemcolor} 0.4\%})&1.00 ({\color{modelcolor} 1.0\%})&6.99 ({\color{problemcolor} 17.3\%})&0.80 ({\color{modelcolor} 16.7\%})\\
CodeLlama-70b-Instruct-hf&0.21&0.93&60.19&1.01&6.76&1.01\\
&0.18 ({\color{problemcolor} 14.3\%})&0.79 ({\color{modelcolor} 15.1\%})&59.49 ({\color{problemcolor} 1.2\%})&1.00 ({\color{modelcolor} 1.0\%})&5.75 ({\color{problemcolor} 14.9\%})&0.86 ({\color{modelcolor} 14.9\%})\\
XwinCoder-13B&0.27&1.08&61.14&1.04&9.25&1.09\\
&0.25 ({\color{problemcolor} 7.4\%})&1.01 ({\color{modelcolor} 6.5\%})&61.15 ({\color{problemcolor} -0.0\%})&1.04 ({\color{modelcolor} 0.0\%})&8.62 ({\color{problemcolor} 6.8\%})&1.02 ({\color{modelcolor} 6.4\%})\\
XwinCoder-34B&0.25&1.07&60.75&1.05&8.46&1.08\\
&0.22 ({\color{problemcolor} 12.0\%})&0.93 ({\color{modelcolor} 13.1\%})&60.75 ({\color{problemcolor} 0.0\%})&1.05 ({\color{modelcolor} 0.0\%})&7.33 ({\color{problemcolor} 13.4\%})&0.94 ({\color{modelcolor} 13.0\%})\\
WizardCoder-7B&0.21&0.91&58.59&1.01&6.63&0.89\\
&0.18 ({\color{problemcolor} 14.3\%})&0.79 ({\color{modelcolor} 13.2\%})&57.97 ({\color{problemcolor} 1.1\%})&1.00 ({\color{modelcolor} 1.0\%})&5.79 ({\color{problemcolor} 12.7\%})&0.78 ({\color{modelcolor} 12.4\%})\\
WizardCoder-13B&0.21&0.81&60.59&1.00&7.22&0.79\\
&0.19 ({\color{problemcolor} 9.5\%})&0.73 ({\color{modelcolor} 9.9\%})&60.53 ({\color{problemcolor} 0.1\%})&1.00 ({\color{modelcolor} 0.0\%})&6.47 ({\color{problemcolor} 10.4\%})&0.71 ({\color{modelcolor} 10.1\%})\\
WizardCoder-34B&0.22&0.79&58.13&1.00&7.10&0.78\\
&0.17 ({\color{problemcolor} 22.7\%})&0.62 ({\color{modelcolor} 21.5\%})&58.42 ({\color{problemcolor} -0.5\%})&1.00 ({\color{modelcolor} 0.0\%})&5.46 ({\color{problemcolor} 23.1\%})&0.60 ({\color{modelcolor} 23.1\%})\\
starcoder2-3b&0.24&1.02&62.45&1.00&7.73&0.89\\
&0.19 ({\color{problemcolor} 20.8\%})&0.79 ({\color{modelcolor} 22.5\%})&62.69 ({\color{problemcolor} -0.4\%})&1.00 ({\color{modelcolor} 0.0\%})&6.68 ({\color{problemcolor} 13.6\%})&0.77 ({\color{modelcolor} 13.5\%})\\
starcoder2-7b&0.21&0.89&62.53&1.00&7.41&0.85\\
&0.19 ({\color{problemcolor} 9.5\%})&0.78 ({\color{modelcolor} 12.4\%})&62.85 ({\color{problemcolor} -0.5\%})&1.00 ({\color{modelcolor} 0.0\%})&6.40 ({\color{problemcolor} 13.6\%})&0.74 ({\color{modelcolor} 12.9\%})\\
        \bottomrule
    \end{tabular}
    \label{tab:humaneval}
\end{table*}

\begin{table*}[ht!]
\tiny
    \centering
    \caption{Evaluation results of \name's effectiveness in the MBPP dataset}
    \begin{tabular}{lrrrrrr}
        \toprule
        Steps&\textbf{ET (s)}&\textbf{NET}&\textbf{MU (Mb)}&\textbf{NMU}&\textbf{TMU (Mb*s)}&\textbf{NTMU}\\
        \midrule
OpenCodeInterpreter-DS-1.3B&0.28&0.94&59.01&1.01&11.73&0.98\\
&0.25 ({\color{problemcolor} 10.7\%})&0.84 ({\color{modelcolor} 10.6\%})&58.99 ({\color{problemcolor} 0.0\%})&1.01 ({\color{modelcolor} 0.0\%})&10.59 ({\color{problemcolor} 9.7\%})&0.89 ({\color{modelcolor} 9.2\%})\\
OpenCodeInterpreter-DS-6.7B&0.26&1.06&58.39&1.00&9.25&1.08\\
&0.21 ({\color{problemcolor} 19.2\%})&0.87 ({\color{modelcolor} 17.9\%})&58.37 ({\color{problemcolor} 0.0\%})&1.00 ({\color{modelcolor} 0.0\%})&7.10 ({\color{problemcolor} 23.2\%})&0.83 ({\color{modelcolor} 23.1\%})\\
OpenCodeInterpreter-DS-33B&0.44&1.59&58.72&1.00&20.19&1.86\\
&0.31 ({\color{problemcolor} 29.5\%})&1.14 ({\color{modelcolor} 28.3\%})&58.70 ({\color{problemcolor} 0.0\%})&1.00 ({\color{modelcolor} 0.0\%})&13.22 ({\color{problemcolor} 34.5\%})&1.22 ({\color{modelcolor} 34.4\%})\\
deepseek-coder-1.3b-instruct&0.63&1.68&354.01&6.05&1463.46&89.12\\
&0.62 ({\color{problemcolor} 1.6\%})&1.64 ({\color{modelcolor} 2.4\%})&339.91 ({\color{problemcolor} 4.0\%})&5.81 ({\color{modelcolor} 4.0\%})&1414.13 ({\color{problemcolor} 3.4\%})&86.11 ({\color{modelcolor} 3.4\%})\\
deepseek-coder-6.7b-instruct&0.76&3.62&58.44&1.00&39.11&5.69\\
&0.21 ({\color{problemcolor} 72.4\%})&0.98 ({\color{modelcolor} 72.9\%})&58.34 ({\color{problemcolor} 0.2\%})&1.00 ({\color{modelcolor} 0.0\%})&6.67 ({\color{problemcolor} 82.9\%})&0.97 ({\color{modelcolor} 83.0\%})\\
deepseek-coder-33b-instruct&0.58&2.33&53.48&0.91&28.74&3.16\\
&0.19 ({\color{problemcolor} 67.2\%})&0.75 ({\color{modelcolor} 67.8\%})&53.34 ({\color{problemcolor} 0.3\%})&0.91 ({\color{modelcolor} 0.0\%})&5.88 ({\color{problemcolor} 79.5\%})&0.65 ({\color{modelcolor} 79.4\%})\\
CodeLlama-7b-Instruct-hf&0.45&2.04&56.96&0.97&13.26&1.79\\
&0.42 ({\color{problemcolor} 6.7\%})&1.89 ({\color{modelcolor} 7.4\%})&56.78 ({\color{problemcolor} 0.3\%})&0.97 ({\color{modelcolor} 0.0\%})&11.98 ({\color{problemcolor} 9.7\%})&1.62 ({\color{modelcolor} 9.5\%})\\
CodeLlama-13b-Instruct-hf&0.53&2.11&55.37&0.95&21.75&2.34\\
&0.52 ({\color{problemcolor} 1.9\%})&2.04 ({\color{modelcolor} 3.3\%})&55.29 ({\color{problemcolor} 0.1\%})&0.95 ({\color{modelcolor} 0.0\%})&21.13 ({\color{problemcolor} 2.9\%})&2.28 ({\color{modelcolor} 2.6\%})\\
CodeLlama-34b-Instruct-hf&0.42&1.18&69.80&1.19&84.01&5.47\\
&0.41 ({\color{problemcolor} 2.4\%})&1.13 ({\color{modelcolor} 4.2\%})&69.32 ({\color{problemcolor} 0.7\%})&1.19 ({\color{modelcolor} 0.0\%})&74.78 ({\color{problemcolor} 11.0\%})&4.87 ({\color{modelcolor} 11.0\%})\\
CodeLlama-70b-Instruct-hf&0.23&1.06&58.13&0.98&7.65&1.05\\
&0.20 ({\color{problemcolor} 13.0\%})&0.93 ({\color{modelcolor} 12.3\%})&58.05 ({\color{problemcolor} 0.1\%})&0.98 ({\color{modelcolor} 0.0\%})&6.67 ({\color{problemcolor} 12.8\%})&0.91 ({\color{modelcolor} 13.3\%})\\
XwinCoder-7B&0.23&1.14&58.45&1.00&7.19&1.10\\
&0.18 ({\color{problemcolor} 21.7\%})&0.90 ({\color{modelcolor} 21.1\%})&58.44 ({\color{problemcolor} 0.0\%})&1.00 ({\color{modelcolor} 0.0\%})&5.89 ({\color{problemcolor} 18.1\%})&0.90 ({\color{modelcolor} 18.2\%})\\
XwinCoder-13B&0.50&1.96&58.38&1.00&23.88&2.50\\
&0.41 ({\color{problemcolor} 18.0\%})&1.61 ({\color{modelcolor} 17.9\%})&58.34 ({\color{problemcolor} 0.1\%})&1.00 ({\color{modelcolor} 0.0\%})&18.95 ({\color{problemcolor} 20.6\%})&1.98 ({\color{modelcolor} 20.8\%})\\
XwinCoder-34B&0.38&1.44&58.27&1.00&14.77&1.48\\
&0.35 ({\color{problemcolor} 7.9\%})&1.32 ({\color{modelcolor} 8.3\%})&58.22 ({\color{problemcolor} 0.1\%})&1.00 ({\color{modelcolor} 0.0\%})&13.54 ({\color{problemcolor} 8.3\%})&1.36 ({\color{modelcolor} 8.1\%})\\
WizardCoder-Python-7B-V1.0-GPTQ&0.22&1.05&58.44&0.99&7.19&1.03\\
&0.20 ({\color{problemcolor} 9.1\%})&0.93 ({\color{modelcolor} 11.4\%})&58.33 ({\color{problemcolor} 0.2\%})&0.99 ({\color{modelcolor} 0.0\%})&6.41 ({\color{problemcolor} 10.8\%})&0.91 ({\color{modelcolor} 11.7\%})\\
WizardCoder-Python-13B-V1.0-GPTQ&0.62&1.35&57.74&0.99&30.66&1.43\\
&0.59 ({\color{problemcolor} 4.8\%})&1.28 ({\color{modelcolor} 5.2\%})&57.70 ({\color{problemcolor} 0.1\%})&0.99 ({\color{modelcolor} 0.0\%})&29.56 ({\color{problemcolor} 3.6\%})&1.38 ({\color{modelcolor} 3.5\%})\\
WizardCoder-Python-34B-V1.0-GPTQ&0.68&2.43&56.75&0.97&34.06&3.14\\
&0.65 ({\color{problemcolor} 4.4\%})&2.33 ({\color{modelcolor} 4.1\%})&56.78 ({\color{problemcolor} -0.1\%})&0.97 ({\color{modelcolor} 0.0\%})&32.63 ({\color{problemcolor} 4.2\%})&3.01 ({\color{modelcolor} 4.1\%})\\
starcoder2-3b&0.17&0.83&45.82&0.79&5.10&0.77\\
&0.16 ({\color{problemcolor} 5.9\%})&0.80 ({\color{modelcolor} 3.6\%})&43.46 ({\color{problemcolor} 5.2\%})&0.74 ({\color{modelcolor} 6.3\%})&4.69 ({\color{problemcolor} 8.0\%})&0.70 ({\color{modelcolor} 9.1\%})\\
starcoder2-7b&1.72&8.63&25.61&0.44&40.42&6.22\\
&1.72 ({\color{problemcolor} 0.0\%})&8.61 ({\color{modelcolor} 0.2\%})&25.56 ({\color{problemcolor} 0.2\%})&0.44 ({\color{modelcolor} 0.0\%})&40.19 ({\color{problemcolor} 0.6\%})&6.19 ({\color{modelcolor} 0.5\%})\\
starcoder2-15b&0.19&1.05&58.62&1.01&6.23&1.05\\
&0.18 ({\color{problemcolor} 5.3\%})&0.99 ({\color{modelcolor} 5.7\%})&58.14 ({\color{problemcolor} 0.8\%})&1.00 ({\color{modelcolor} 1.0\%})&5.92 ({\color{problemcolor} 5.0\%})&1.00 ({\color{modelcolor} 4.8\%})\\
        \bottomrule
    \end{tabular}
    \label{tab:mbpp}
\end{table*}





\subsection{Detailed Evaluation Metric for Efficiency}
\label{app:detailed-metric}

In our work, we adopt the efficiency metrics proposed by EffiBench~\citep{huang2024effibench} to evaluate the effectiveness of \name in improving the efficiency of LLM-generated code.

\paragraph{Execution Time (ET)} Execution time (ET) measures the average time taken for code execution. Mathematically, ET is defined as:
$$ET = \frac{1}{N}\sum^{N}T_{\text{code}}$$
where $ET$ is the execution time metric, $T_{\text{code}}$ is the execution time of the code (with all the test cases), and $N$ is the number of codes generated by code generation models used for evaluation. 

\paragraph{Normalized Execution Time (NET)} Normalized Execution Time (NET) measures the execution time required by generated code relative to that of a canonical solution. We define NET as:
$$NET = \frac{1}{N}\sum^{N}\frac{T_{\text{code}}}{T_{\text{canonical}}}$$
where $T_{\text{code}}$ is the execution time of the generated code, and $T_{\text{canonical}}$ is the execution time of the canonical solution. A NET value greater than 1 indicates that the generated code is slower than the canonical solution, while a value less than 1 suggests the generated code is faster.

\paragraph{Max Memory Usage (MU)} Max Memory Usage (MU) measures the average max memory consumption during code execution. Mathematically, MU is defined as:
$$MU = \frac{1}{N}\sum^{N}M_{\text{code}}$$
where $MU$ is the memory usage metric, $M_{\text{code}}$ is the max memory consumption of the generated code among all the test cases, and $N$ is the number of code instances generated by code generation models used for evaluation. This metric is critical for assessing the resource efficiency of generated code, particularly in environments with limited maximum memory capacity.

\paragraph{Normalized Max Memory Usage (NMU)} Normalized Max Memory Usage (NMU) quantifies how the max memory efficiency of the generated code compares to the canonical solution. We define NMU as:
$$NMU = \frac{1}{N}\sum^{N}\frac{M_{\text{code}}}{M_{\text{canonical}}}$$
where $NMU$ is the normalized max memory usage metric, $M_{\text{code}}$ is the max memory usage of the generated code, and $M_{\text{canonical}}$ is the max memory usage of the canonical solution. An NMU value less than 1 indicates that the generated code is more memory-efficient than the canonical solution, whereas a value greater than 1 suggests it is less efficient in terms of memory usage. This metric provides a relative measure of the memory optimization in the generated code in comparison to a standard baseline.

\paragraph{Total Memory Usage (TMU)} Total Memory Usage (TMU) assesses the efficiency of memory usage throughout the execution of code, taking into account both the magnitude and duration of memory utilization. To calculate TMU, first, monitor and record the memory usage at discrete time intervals during the execution, resulting in a memory usage profile $M(t)$, where $t$ represents time. Then, compute the area under the curve of $M(t)$ over the total execution time, $T_{\text{total}}$, using numerical integration methods such as the trapezoidal rule:
$$TMU = \frac{1}{N}\sum^{N}\int_0^{T_{\text{total}}} M(t) \, dt$$
A lower TMU value indicates higher memory efficiency, reflecting an optimized balance between the amount of memory used and the duration of its usage.

\paragraph{Normalized Total Memory Usage (NTMU)} The Normalized Total Memory Usage (NTMU) offers a comparison of the dynamic memory efficiency between the generated code and the canonical solution. To determine NTMU, calculate the TMU for both the generated code and the canonical solution. Normalize the TMU of the generated code by dividing it by the TMU of the canonical solution:
$$NTMU = \frac{1}{N}\sum^{N}\frac{TMU_{\text{code}}}{TMU_{\text{canonical}}}$$
where \(TMU_{\text{code}}\) is the TMU of the generated code and \(TMU_{\text{canonical}}\) is the TMU of the canonical solution. An NTMU value less than 1 signifies that the generated code manages dynamic memory more efficiently compared to the canonical solution, while a value greater than 1 indicates a less efficient management of dynamic memory. This metric provides insight into the relative use of dynamic memory of generated code compared to an established benchmark.

\subsection{Additional Related Work}
\paragraph{Instruction Tuning for Code} Instruction tuning has proven effective in enhancing the usability and overall performance of LLMs across various language tasks~\citep{Ouyang0JAWMZASR22,WeiBZGYLDDL22,viswanathan2023prompt2model,zhao2024self}. This approach has been extended to the domain of code generation. The core challenge is the acquisition of high-quality instructional data, which is often labor-intensive. To address this, recent research has focused on developing methods to generate synthetic instruction data. Studies have shown that textbook-quality synthetic data alone can improve a model's coding and reasoning capabilities~\citep{Gunasekar2023, Liphi2023}. One early effort was Self-Instruct~\citep{WangKMLSKH23}, which utilized LLMs to generate synthetic instruction-response pairs using carefully crafted prompts. The same LLM was then instruction-tuned on this synthetic data. Code Alpaca~\citep{Codealpaca} applied the Self-Instruct approach with GPT models, tailoring it specifically for code generation, editing, and optimization tasks. Building upon this, WizardCoder~\citep{WizardCoder2024} adapted the Evol-Instruct technique~\citep{WizardLM2024} to the coding domain by designing heuristic prompts to create more complex and diverse synthetic data. OSS-Instruct~\citep{Magicoder2024} took a different approach by leveraging LLMs to automatically generate new coding problems inspired by random code snippets from open-source repositories. In contrast, Octopack~\citep{OctoPack2024} focused on collecting and filtering high-quality Git commit messages that resemble natural language instructions. 

\begin{figure*}[htbp]
    \centering
    \begin{minipage}{1\textwidth}
        \begin{problembox}
            \textbf{Problem:} You have a lock in front of you with 4 circular wheels. Each wheel has 10 slots: '0', '1', '2', '3', '4', '5', '6', '7', '8', '9'. The wheels can rotate freely and wrap around: for example we can turn '9' to be '0', or '0' to be '9'. Each move consists of turning one wheel one slot.
            
            The lock initially starts at '0000', a string representing the state of the 4 wheels.
            
            You are given a list of deadends dead ends, meaning if the lock displays any of these codes, the wheels of the lock will stop turning and you will be unable to open it.
            
            Given a target representing the value of the wheels that will unlock the lock, return the minimum total number of turns required to open the lock, or -1 if it is impossible.
            
            \begin{description}[leftmargin=4pt]
                \item[Example 1:] \hfill \\
                \textbf{Input:} deadends = ["0201","0101","0102","1212","2002"], target = "0202"\\
                \textbf{Output:} 6\\
                \textbf{Explanation:} A sequence of valid moves would be "0000" $->$ "1000" $->$ "1100" $->$ "1200" $->$ "1201" $->$ "1202" $->$ "0202". Note that a sequence like "0000" $->$ "0001" $->$ "0002" $->$ "0102" $->$ "0202" would be invalid, because the wheels of the lock become stuck after the display becomes the dead end "0102".
                
                \item[Example 2:] \hfill \\
                \textbf{Input:} deadends = ["8888"], target = "0009"\\
                \textbf{Output:} 1\\
                \textbf{Explanation:} We can turn the last wheel in reverse to move from "0000" $->$ "0009".
                
                \item[Example 3:] \hfill \\
                \textbf{Input:} deadends = ["8887","8889","8878","8898","8788","8988","7888","9888"], target = "8888"\\
                \textbf{Output:} -1\\
                \textbf{Explanation:} We cannot reach the target without getting stuck.
            \end{description}
            \textbf{Constraints:}
            \begin{itemize}
                \item 1 $\leq$ deadends.length $\leq$ 500
                \item deadends[i].length == 4
                \item target.length == 4
                \item target will not be in the list deadends.
                \item target and deadends[i] consist of digits only.
            \end{itemize}
\begin{minted}{python}
solution = Solution()
assert solution.openLock(["0201","0101","0102","1212","2002"], "0202") == 6
assert solution.openLock(["8888"], "0009") == 1
assert solution.openLock(["8887","8889","8878","8898","8788","8988","7888","9888"], "8888") == -1
\end{minted}
        \end{problembox}
    \end{minipage}
    \caption{A case illustration of GPT-4-turbo-preview (OpenLock). Part A: Task Description.}
    \label{fig:effective_case_task_description}
\end{figure*}

\begin{figure*}[htbp]
    \centering
        \begin{minipage}[t]{1\textwidth}
            \begin{modelbox}[title=\textbf{Completion}]
\begin{minted}{python}
from typing import List
from collections import deque

class Solution:
    def openLock(self, deadends: List[str], target: str) -> int:
        def neighbors(node):
            for i in range(4):
                x = int(node[i])
                for d in (-1, 1):
                    y = (x + d) % 10
                    yield node[:i] + str(y) + node[i+1:]
        
        dead = set(deadends)
        queue = deque([('0000', 0)])
        seen = {'0000'}
        
        while queue:
            node, depth = queue.popleft()
            if node == target: 
                return depth
            if node in dead: 
                continue
            for nei in neighbors(node):
                if nei not in seen:
                    seen.add(nei)
                    queue.append((nei, depth + 1))
        
        return -1
\end{minted}
            \end{modelbox}
        \end{minipage}%
    \caption{A case illustration of GPT-4-turbo-preview (OpenLock). Part B: GPT-4-turbo-preview generated code.}
    \label{fig:effective_case_task_code}
\end{figure*}

\begin{figure*}[htbp]
    \centering
        \begin{minipage}[t]{1\textwidth}
            \begin{modelbox}[title=\textbf{Execution Time Profile}]
\begin{minted}{python}
Total time: 23.5891 s
File: 752_tmp.py
Function: openLock at line 24

Line #      Hits         Time  Per Hit   % Time  Line Contents
==============================================================
    24                                               @profile
    25                                               def openLock(self, deadends: List[str], target: str) -> int:
    26       200      14151.9     70.8      0.1          @profile
    27       200        107.6      0.5      0.0          def neighbors(node):
    28                                                       for i in range(4):
    29                                                           x = int(node[i])
    30                                                           for d in (-1, 1):
    31                                                               y = (x + d) % 10
    32                                                               yield node[:i] + str(y) + node[i+1:]
    33                                                   
    34       100        189.1      1.9      0.0          dead = set(deadends)
    35       100        125.8      1.3      0.0          queue = deque([('0000', 0)])
    36       100         58.8      0.6      0.0          seen = {'0000'}
    37                                                   
    38    478172     109953.5      0.2      0.5          while queue:
    39    478172     183034.8      0.4      0.8              node, depth = queue.popleft()
    40    478172     113871.5      0.2      0.5              if node == target: 
    41       100         18.8      0.2      0.0                  return depth
    42    478072     105351.0      0.2      0.4              if node in dead: 
    43       253         44.3      0.2      0.0                  continue
    44   4300371   21651001.1      5.0     91.8              for nei in neighbors(node):
    45   3822552     969789.3      0.3      4.1                  if nei not in seen:
    46    576544     200262.9      0.3      0.8                      seen.add(nei)
    47    576544     241153.3      0.4      1.0                      queue.append((nei, depth + 1))
    48                                                   
    49                                                   return -1

Total time: 0.0129757 s
File: 752_tmp.py
Function: neighbors at line 26

Line #      Hits         Time  Per Hit   % Time  Line Contents
==============================================================
    26                                                   @profile
    27                                                   def neighbors(node):
    28      5345       1206.2      0.2      9.3              for i in range(4):
    29      4276       1595.9      0.4     12.3                  x = int(node[i])
    30     12828       2305.9      0.2     17.8                  for d in (-1, 1):
    31      8552       2248.4      0.3     17.3                      y = (x + d) % 10
    32      8552       5619.3      0.7     43.3                      yield node[:i] + str(y) + node[i+1:]
\end{minted}
            \end{modelbox}
        \end{minipage}%
    \caption{A case illustration of GPT-4-turbo-preview (OpenLock). Part C: Execution time profile.}
    \label{fig:effective_case_task_time}
\end{figure*}

\begin{figure*}[htbp]
    \centering
        \begin{minipage}[t]{1\textwidth}
            \begin{modelbox}[title=\textbf{Memory Usage Profile}]
\begin{minted}{python}
Line #    Mem usage    Increment  Occurrences   Line Contents
=============================================================
    14     48.5 MiB     48.5 MiB           1       @profile(stream=profile_stream, precision=PROFILE_PRECISION)
    15                                             def openLock(self, deadends: List[str], target: str) -> int:
    16     48.8 MiB      0.0 MiB        4225           def neighbors(node):
    17     48.8 MiB      0.0 MiB       21123               for i in range(4):
    18     48.8 MiB      0.0 MiB       16898                   x = int(node[i])
    19     48.8 MiB      0.0 MiB       50696                   for d in (-1, 1):
    20     48.8 MiB      0.0 MiB       33797                       y = (x + d) % 10
    21     48.8 MiB      0.1 MiB       67594                       yield node[:i] + str(y) + node[i+1:]
    22                                                 
    23     48.5 MiB      0.0 MiB           1           dead = set(deadends)
    24     48.5 MiB      0.0 MiB           1           queue = deque([('0000', 0)])
    25     48.5 MiB      0.0 MiB           1           seen = {'0000'}
    26                                                 
    27     48.8 MiB      0.0 MiB        4229           while queue:
    28     48.8 MiB      0.1 MiB        4229               node, depth = queue.popleft()
    29     48.8 MiB      0.0 MiB        4229               if node == target: 
    30     48.8 MiB      0.0 MiB           1                   return depth
    31     48.8 MiB      0.0 MiB        4228               if node in dead: 
    32                                                         continue
    33     48.8 MiB      0.0 MiB       38022               for nei in neighbors(node):
    34     48.8 MiB      0.0 MiB       33797                   if nei not in seen:
    35     48.8 MiB      0.2 MiB        4977                       seen.add(nei)
    36     48.8 MiB      0.0 MiB        4977                       queue.append((nei, depth + 1))
    37                                                 
    38                                                 return -1
\end{minted}
            \end{modelbox}
        \end{minipage}%
    \caption{A case illustration of GPT-4-turbo-preview (OpenLock). Part D: Memory usage profile.}
    \label{fig:effective_case_task_memory}
\end{figure*}

\begin{figure*}[htbp]
    \centering
        \begin{minipage}[t]{1\textwidth}
            \begin{modelbox}[title=\textbf{\name Completion}]
\begin{minted}{python}
from typing import List
from collections import deque

class Solution:
    def openLock(self, deadends: List[str], target: str) -> int:
        """Calculate the minimum number of moves to open the lock from '0000' to the target avoiding deadends."""
        if '0000' in deadends:
            return -1
        if target == '0000':
            return 0
        
        def getNextStates(state: str) -> List[str]:
            """Generate the next states for a given lock state."""
            res = []
            for i in range(4):
                current_digit = int(state[i])
                for diff in [-1, 1]:  # decrease or increase the digit
                    new_digit = (current_digit + diff) % 10
                    res.append(state[:i] + str(new_digit) + state[i+1:])
            return res

        def bfs(queue: Deque[str], distances: dict, other_distances: dict) -> int:
            """Perform a step of BFS, expanding the current queue."""
            for _ in range(len(queue)):
                current = queue.popleft()
                step_count = distances[current]
                for nextState in getNextStates(current):
                    if nextState in deadends_set or nextState in distances:
                        continue
                    if nextState in other_distances:
                        return step_count + 1 + other_distances[nextState]
                    distances[nextState] = step_count + 1
                    queue.append(nextState)
            return -1

        deadends_set = set(deadends)
        # Early exit if the initial state is a deadend
        if '0000' in deadends_set:
            return -1
        
        # Initialize BFS structures for bidirectional search
        queue1, queue2 = deque(['0000']), deque([target])
        distance1, distance2 = {"0000": 0}, {target: 0}
        
        # Perform BFS from both ends
        while queue1 and queue2:
            result = bfs(queue1, distance1, distance2)
            if result != -1:
                return result
            result = bfs(queue2, distance2, distance1)
            if result != -1:
                return result
        
        return -1
\end{minted}
            \end{modelbox}
    \end{minipage}
    \caption{A case illustration of GPT-4-turbo-preview (OpenLock). Part E: Self-Optimization Code.}
    \label{fig:effective_case_task_refine_code}
\end{figure*}

\begin{figure*}[htbp]
    \centering
    \begin{minipage}[t]{1\textwidth}
        \begin{modelbox}[title=\textbf{Execution Time Profile}]
\begin{minted}{python}
Total time: 3.35559 s
File: 752_tmp.py
Function: openLock at line 24

Line #      Hits         Time  Per Hit   % Time  Line Contents
==============================================================
    24                                               @profile
    25                                               def openLock(self, deadends: List[str], target: str) -> int:
    26                                                   """Calculate the minimum number of moves to open the lock from '0000' to the target avoiding deadends."""
    27       100         65.1      0.7      0.0          if '0000' in deadends:
    28                                                       return -1
    29       100         35.0      0.3      0.0          if target == '0000':
    30                                                       return 0
    31                                                   

    57                                           
    58       100         99.3      1.0      0.0          deadends_set = set(deadends)
    59                                                   # Early exit if the initial state is a deadend
    60       100         36.7      0.4      0.0          if '0000' in deadends_set:
    61                                                       return -1
    62                                                   
    63                                                   # Initialize BFS structures for bidirectional search
    64       100        116.8      1.2      0.0          queue1, queue2 = deque(['0000']), deque([target])
    65       100         47.6      0.5      0.0          distance1, distance2 = {"0000": 0}, {target: 0}
    66                                                   
    67                                                   # Perform BFS from both ends
    68       518        154.2      0.3      0.0          while queue1 and queue2:
    69       518    1936061.9   3737.6     57.7              result = bfs(queue1, distance1, distance2)
    70       518        153.0      0.3      0.0              if result != -1:
    71        52         10.3      0.2      0.0                  return result
    72       466    1388682.7   2980.0     41.4              result = bfs(queue2, distance2, distance1)
    73       466        136.6      0.3      0.0              if result != -1:
    74        48          9.5      0.2      0.0                  return result
    75                                                   
    76                                                   return -1
\end{minted}
        \end{modelbox}
    \end{minipage}%
    \caption{A case illustration of GPT-4-turbo-preview (OpenLock). Part F1: Execution time profile for Self-Optimization Code.}
    \label{fig:effective_case_task_refine_code_time}
\end{figure*}

\begin{figure*}[htbp]
    \centering
    \begin{minipage}[t]{1\textwidth}
        \begin{modelbox}[title=\textbf{Execution Time Profile}]
\begin{minted}{python}

Total time: 0.000697948 s
File: 752_tmp.py
Function: getNextStates at line 32

Line #      Hits         Time  Per Hit   % Time  Line Contents
==============================================================
    32                                                   @profile
    33                                                   def getNextStates(state: str) -> List[str]:
    34                                                       """Generate the next states for a given lock state."""
    35        52         12.1      0.2      1.7              res = []
    36       260         62.4      0.2      8.9              for i in range(4):
    37       208         74.4      0.4     10.7                  current_digit = int(state[i])
    38       624        127.4      0.2     18.2                  for diff in [-1, 1]:  # decrease or increase the digit
    39       416        104.5      0.3     15.0                      new_digit = (current_digit + diff) % 10
    40       416        308.4      0.7     44.2                      res.append(state[:i] + str(new_digit) + state[i+1:])
    41        52          8.8      0.2      1.3              return res

Total time: 0.00182689 s
File: 752_tmp.py
Function: bfs at line 43

Line #      Hits         Time  Per Hit   % Time  Line Contents
==============================================================
    43                                                   @profile
    44                                                   def bfs(queue: Deque[str], distances: dict, other_distances: dict) -> int:
    45                                                       """Perform a step of BFS, expanding the current queue."""
    46        57         12.9      0.2      0.7              for _ in range(len(queue)):
    47        52         14.0      0.3      0.8                  current = queue.popleft()
    48        52         11.4      0.2      0.6                  step_count = distances[current]
    49       462       1499.4      3.2     82.1                  for nextState in getNextStates(current):
    50       411        114.9      0.3      6.3                      if nextState in deadends_set or nextState in distances:
    51       235         34.5      0.1      1.9                          continue
    52       176         38.7      0.2      2.1                      if nextState in other_distances:
    53         1          0.7      0.7      0.0                          return step_count + 1 + other_distances[nextState]
    54       175         47.5      0.3      2.6                      distances[nextState] = step_count + 1
    55       175         51.9      0.3      2.8                      queue.append(nextState)
    56         5          0.9      0.2      0.0              return -1
\end{minted}
        \end{modelbox}
    \end{minipage}%
    \caption{A case illustration of GPT-4-turbo-preview (OpenLock). Part F2: Execution time profile for Self-Optimization Code.}
    \label{fig:effective_case_task_refine_code_time}
\end{figure*}

\begin{figure*}[htbp]
    \centering
        \begin{minipage}[t]{1\textwidth}
            \begin{modelbox}[title=\textbf{Memory Usage Profile}]
\begin{minted}{python}
Line #    Mem usage    Increment  Occurrences   Line Contents
=============================================================
    14     48.4 MiB     48.4 MiB           1       @profile(stream=profile_stream, precision=PROFILE_PRECISION)
    15                                             def openLock(self, deadends: List[str], target: str) -> int:
    16                                                 """Calculate the minimum number of moves to open the lock from '0000' to the target avoiding deadends."""
    17     48.4 MiB      0.0 MiB           1           if '0000' in deadends:
    18                                                     return -1
    19     48.4 MiB      0.0 MiB           1           if target == '0000':
    20                                                     return 0
    21                                                 
    22     48.4 MiB      0.0 MiB         819           def getNextStates(state: str) -> List[str]:
    23                                                     """Generate the next states for a given lock state."""
    24     48.4 MiB      0.0 MiB         818               res = []
    25     48.4 MiB      0.0 MiB        4090               for i in range(4):
    26     48.4 MiB      0.0 MiB        3272                   current_digit = int(state[i])
    27     48.4 MiB      0.0 MiB        9816                   for diff in [-1, 1]:  # decrease or increase the digit
    28     48.4 MiB      0.0 MiB        6544                       new_digit = (current_digit + diff) % 10
    29     48.4 MiB      0.0 MiB        6544                       res.append(state[:i] + str(new_digit) + state[i+1:])
    30     48.4 MiB      0.0 MiB         818               return res
    31                                         
    32     48.4 MiB      0.0 MiB          10           def bfs(queue: Deque[str], distances: dict, other_distances: dict) -> int:
    33                                                     """Perform a step of BFS, expanding the current queue."""
    34     48.4 MiB      0.0 MiB         826               for _ in range(len(queue)):
    35     48.4 MiB      0.0 MiB         818                   current = queue.popleft()
    36     48.4 MiB      0.0 MiB         818                   step_count = distances[current]
    37     48.4 MiB      0.0 MiB        7356                   for nextState in getNextStates(current):
    38     48.4 MiB      0.0 MiB        6539                       if nextState in deadends_set or nextState in distances:
    39     48.4 MiB      0.0 MiB        5057                           continue
    40     48.4 MiB      0.0 MiB        1482                       if nextState in other_distances:
    41     48.4 MiB      0.0 MiB           1                           return step_count + 1 + other_distances[nextState]
    42     48.4 MiB      0.1 MiB        1481                       distances[nextState] = step_count + 1
    43     48.4 MiB      0.0 MiB        1481                       queue.append(nextState)
    44     48.4 MiB      0.0 MiB           8               return -1
    45                                         
    46     48.4 MiB      0.0 MiB           1           deadends_set = set(deadends)
    47                                                 # Early exit if the initial state is a deadend
    48     48.4 MiB      0.0 MiB           1           if '0000' in deadends_set:
    49                                                     return -1
    50                                                 
    51                                                 # Initialize BFS structures for bidirectional search
    52     48.4 MiB      0.0 MiB           1           queue1, queue2 = deque(['0000']), deque([target])
    53     48.4 MiB      0.0 MiB           1           distance1, distance2 = {"0000": 0}, {target: 0}
    54                                                 
    55                                                 # Perform BFS from both ends
    56     48.4 MiB      0.0 MiB           5           while queue1 and queue2:
    57     48.4 MiB      0.0 MiB           5               result = bfs(queue1, distance1, distance2)
    58     48.4 MiB      0.0 MiB           5               if result != -1:
    59                                                         return result
    60     48.4 MiB      0.0 MiB           4               result = bfs(queue2, distance2, distance1)
    61     48.4 MiB      0.0 MiB           4               if result != -1:
    62                                                         return result
    63                                                 
    64                                                 return -1
\end{minted}
            \end{modelbox}
        \end{minipage}%
    \caption{A case illustration of GPT-4-turbo-preview (OpenLock). Part G: Memory usage profile.}
    \label{fig:effective_case_task_refine_code_memory}
\end{figure*}

\begin{figure*}[htbp]
    \centering
    \begin{minipage}{1\textwidth}
    \begin{problembox}
        \textbf{Problem:} Given two sorted arrays \texttt{nums1} and \texttt{nums2} of size \texttt{m} and \texttt{n} respectively, return \textbf{the median} of the two sorted arrays.

        The overall run time complexity should be \(\mathcal{O}(\log (m+n))\).

        \textbf{Example 1:}
        \begin{verbatim}
        Input: nums1 = [1,3], nums2 = [2]
        Output: 2.00000
        Explanation: merged array = [1,2,3] and median is 2.
        \end{verbatim}

        \textbf{Example 2:}
        \begin{verbatim}
        Input: nums1 = [1,2], nums2 = [3,4]
        Output: 2.50000
        Explanation: merged array = [1,2,3,4] 
        and median is (2 + 3) / 2 = 2.5.
        \end{verbatim}

        \textbf{Constraints:}
        \begin{itemize}
            \item \texttt{nums1.length == m}
            \item \texttt{nums2.length == n}
            \item \(0 \leq m \leq 1000\)
            \item \(0 \leq n \leq 1000\)
            \item \(1 \leq m + n \leq 2000\)
            \item \(-10^6 \leq \texttt{nums1[i]}, \texttt{nums2[i]} \leq 10^6\)
        \end{itemize}
\begin{minted}{python}
solution = Solution()
assert solution.findMedianSortedArrays([1, 3], [2]) == 2.0
assert solution.findMedianSortedArrays([1, 2], [3, 4]) == 2.5
\end{minted}
    \end{problembox}
    \end{minipage}
    \caption{A case illustration of GPT-4-turbo-preview (FindMedianSortedArrays). Part A: Task Description.}
    \label{fig:ineffective_case_task_description}
\end{figure*}

\begin{figure*}[htbp]
    \centering
        \begin{minipage}[t]{1\textwidth}
            \begin{modelbox}[title=\textbf{Task Description}]
\begin{minted}{python}
class Solution:
    def findMedianSortedArrays(self, nums1, nums2) -> float:
        A, B = nums1, nums2
        total = len(A) + len(B)
        half = total // 2
        
        if len(B) < len(A):
            A, B = B, A
            
        l, r = 0, len(A) - 1
        while True:
            i = (l + r) // 2 # A
            j = half - i - 2 # B
            
            Aleft = A[i] if i >= 0 else float("-infinity")
            Aright = A[i + 1] if (i + 1) < len(A) else float("infinity")
            Bleft = B[j] if j >= 0 else float("-infinity")
            Bright = B[j + 1] if (j + 1) < len(B) else float("infinity")
            
            if Aleft <= Bright and Bleft <= Aright:
                if total % 2:
                    return min(Aright, Bright)
                return (max(Aleft, Bleft) + min(Aright, Bright)) / 2
            elif Aleft > Bright:
                r = i - 1
            else:
                l = i + 1
\end{minted}
            \end{modelbox}
        \end{minipage}%
    \caption{A case illustration of GPT-4-turbo-preview (FindMedianSortedArrays). Part B: GPT-4-turbo-preview generated code.}
    \label{fig:ineffective_case_task_code}
\end{figure*}

\begin{figure*}[htbp]
    \centering
        \begin{minipage}[t]{1\textwidth}
            \begin{modelbox}[title=\textbf{Task Description}]
\begin{minted}{python}
Timer unit: 1e-06 s

Total time: 0.00119913 s
File: 4_tmp.py
Function: findMedianSortedArrays at line 45

Line #      Hits         Time  Per Hit   % Time  Line Contents
==============================================================
    45                                               @profile
    46                                               def findMedianSortedArrays(self, nums1, nums2) -> float:
    47       100         41.2      0.4      3.4          A, B = nums1, nums2
    48       100         57.2      0.6      4.8          total = len(A) + len(B)
    49       100         40.6      0.4      3.4          half = total // 2
    50                                                   
    51       100         54.9      0.5      4.6          if len(B) < len(A):
    52        46         15.9      0.3      1.3              A, B = B, A
    53                                                       
    54       100         42.9      0.4      3.6          l, r = 0, len(A) - 1
    55       169         58.6      0.3      4.9          while True:
    56       169         63.3      0.4      5.3              i = (l + r) // 2 # A
    57       169         62.1      0.4      5.2              j = half - i - 2 # B
    58                                                       
    59       169         83.8      0.5      7.0              Aleft = A[i] if i >= 0 else float("-infinity")
    60       169         99.3      0.6      8.3              Aright = A[i + 1] if (i + 1) < len(A) else float("infinity")
    61       169         84.3      0.5      7.0              Bleft = B[j] if j >= 0 else float("-infinity")
    62       169         95.3      0.6      8.0              Bright = B[j + 1] if (j + 1) < len(B) else float("infinity")
    63                                                       
    64       169         98.0      0.6      8.2              if Aleft <= Bright and Bleft <= Aright:
    65       100         51.6      0.5      4.3                  if total % 2:
    66        51         86.1      1.7      7.2                      return min(Aright, Bright)
    67        49        103.0      2.1      8.6                  return (max(Aleft, Bleft) + min(Aright, Bright)) / 2
    68        69         36.9      0.5      3.1              elif Aleft > Bright:
    69        45         15.8      0.4      1.3                  r = i - 1
    70                                                       else:
    71        24          8.4      0.3      0.7                  l = i + 1

\end{minted}
            \end{modelbox}
        \end{minipage}%
    \caption{A case illustration of GPT-4-turbo-preview (FindMedianSortedArrays). Part C: Execution time profile.}
    \label{fig:ineffective_case_task_code_time}
\end{figure*}

\begin{figure*}[htbp]
    \centering
        \begin{minipage}[t]{1\textwidth}
            \begin{modelbox}[title=\textbf{Task Description}]
\begin{minted}{python}
Line #    Mem usage    Increment  Occurrences   Line Contents
=============================================================
    54     24.7 MiB     24.7 MiB           1       @profile(stream=profile_stream, precision=PROFILE_PRECISION)
    55                                             def findMedianSortedArrays(self, nums1, nums2) -> float:
    56     24.7 MiB      0.0 MiB           1           A, B = nums1, nums2
    57     24.7 MiB      0.0 MiB           1           total = len(A) + len(B)
    58     24.7 MiB      0.0 MiB           1           half = total // 2
    59                                                 
    60     24.7 MiB      0.0 MiB           1           if len(B) < len(A):
    61                                                     A, B = B, A
    62                                                     
    63     24.7 MiB      0.0 MiB           1           l, r = 0, len(A) - 1
    64     24.7 MiB      0.0 MiB           1           while True:
    65     24.7 MiB      0.0 MiB           1               i = (l + r) // 2 # A
    66     24.7 MiB      0.0 MiB           1               j = half - i - 2 # B
    67                                                     
    68     24.7 MiB      0.0 MiB           1               Aleft = A[i] if i >= 0 else float("-infinity")
    69     24.7 MiB      0.0 MiB           1               Aright = A[i + 1] if (i + 1) < len(A) else float("infinity")
    70     24.7 MiB      0.0 MiB           1               Bleft = B[j] if j >= 0 else float("-infinity")
    71     24.7 MiB      0.0 MiB           1               Bright = B[j + 1] if (j + 1) < len(B) else float("infinity")
    72                                                     
    73     24.7 MiB      0.0 MiB           1               if Aleft <= Bright and Bleft <= Aright:
    74     24.7 MiB      0.0 MiB           1                   if total % 2:
    75                                                             return min(Aright, Bright)
    76                                                         return (max(Aleft, Bleft) + min(Aright, Bright)) / 2
    77                                                     elif Aleft > Bright:
    78                                                         r = i - 1
    79                                                     else:
    80                                                         l = i + 1
\end{minted}
            \end{modelbox}
        \end{minipage}%
    \caption{A case illustration of GPT-4-turbo-preview (FindMedianSortedArrays). Part D: Memory usage profile.}
    \label{fig:ineffective_case_task_code_memory}
\end{figure*}

\begin{figure*}[htbp]
    \centering
        \begin{minipage}[t]{1\textwidth}
            \begin{modelbox}[title=\textbf{Task Description}]
\begin{minted}{python}
class Solution:
    def findMedianSortedArrays(self, nums1, nums2) -> float:
        # Ensure nums1 is the smaller array to minimize binary search domain
        if len(nums1) > len(nums2):
            nums1, nums2 = nums2, nums1
        x, y = len(nums1), len(nums2)

        low, high = 0, x
        while low <= high:
            partitionX = (low + high) // 2
            partitionY = (x + y + 1) // 2 - partitionX

            # If partitionX is 0 it means nothing is there on left side. Use -inf for maxLeftX
            maxLeftX = float('-inf') if partitionX == 0 else nums1[partitionX - 1]
            # If partitionX is length of input then there is nothing on right side. Use +inf for minRightX
            minRightX = float('inf') if partitionX == x else nums1[partitionX]

            maxLeftY = float('-inf') if partitionY == 0 else nums2[partitionY - 1]
            minRightY = float('inf') if partitionY == y else nums2[partitionY]

            if maxLeftX <= minRightY and maxLeftY <= minRightX:
                # The partition is correct
                if (x + y) % 2 == 0:
                    return (max(maxLeftX, maxLeftY) + min(minRightX, minRightY)) / 2.0
                else:
                    return max(maxLeftX, maxLeftY)
            elif maxLeftX > minRightY:  # We are too far on right side for partitionX. Go on left side.
                high = partitionX - 1
            else:  # We are too far on left side for partitionX. Go on right side.
                low = partitionX + 1
\end{minted}
            \end{modelbox}
        \end{minipage}%
    \caption{A case illustration of GPT-4-turbo-preview (FindMedianSortedArrays). Part E: Self-Optimization Code.}
    \label{fig:ineffective_case_task_code_refine_code}
\end{figure*}

\begin{figure*}[htbp]
    \centering
        \begin{minipage}[t]{1\textwidth}
            \begin{modelbox}[title=\textbf{Task Description}]
\begin{minted}{python}
Timer unit: 1e-06 s

Total time: 0.00114754 s
File: 4_tmp.py
Function: findMedianSortedArrays at line 45

Line #      Hits         Time  Per Hit   % Time  Line Contents
==============================================================
    45                                               @profile
    46                                               def findMedianSortedArrays(self, nums1, nums2) -> float:
    47                                                   # Ensure nums1 is the smaller array to minimize binary search domain
    48       100         63.3      0.6      5.5          if len(nums1) > len(nums2):
    49        46         18.2      0.4      1.6              nums1, nums2 = nums2, nums1
    50       100         44.7      0.4      3.9          x, y = len(nums1), len(nums2)
    51                                           
    52       100         32.3      0.3      2.8          low, high = 0, x
    53       184         79.6      0.4      6.9          while low <= high:
    54       184         72.1      0.4      6.3              partitionX = (low + high) // 2
    55       184         69.7      0.4      6.1              partitionY = (x + y + 1) // 2 - partitionX
    56                                           
    57                                                       # If partitionX is 0 it means nothing is there on left side. Use -inf for maxLeftX
    58       184         92.6      0.5      8.1              maxLeftX = float('-inf') if partitionX == 0 else nums1[partitionX - 1]
    59                                                       # If partitionX is length of input then there is nothing on right side. Use +inf for minRightX
    60       184         81.5      0.4      7.1              minRightX = float('inf') if partitionX == x else nums1[partitionX]
    61                                           
    62       184         94.9      0.5      8.3              maxLeftY = float('-inf') if partitionY == 0 else nums2[partitionY - 1]
    63       184         86.9      0.5      7.6              minRightY = float('inf') if partitionY == y else nums2[partitionY]
    64                                           
    65       184        100.1      0.5      8.7              if maxLeftX <= minRightY and maxLeftY <= minRightX:
    66                                                           # The partition is correct
    67       100         52.8      0.5      4.6                  if (x + y) % 2 == 0:
    68        49        108.3      2.2      9.4                      return (max(maxLeftX, maxLeftY) + min(minRightX, minRightY)) / 2.0
    69                                                           else:
    70        51         86.7      1.7      7.6                      return max(maxLeftX, maxLeftY)
    71        84         34.7      0.4      3.0              elif maxLeftX > minRightY:  # We are too far on right side for partitionX. Go on left side.
    72        30         11.0      0.4      1.0                  high = partitionX - 1
    73                                                       else:  # We are too far on left side for partitionX. Go on right side.
    74        54         18.2      0.3      1.6                  low = partitionX + 1

\end{minted}
            \end{modelbox}
        \end{minipage}%
    \caption{A case illustration of GPT-4-turbo-preview (FindMedianSortedArrays). Part F: Execution time profile.}
    \label{fig:ineffective_case_task_code_refine_code_time}
\end{figure*}

\begin{figure*}[htbp]
    \centering
        \begin{minipage}[t]{1\textwidth}
            \begin{modelbox}[title=\textbf{Task Description}]
\begin{minted}{python}
Line #    Mem usage    Increment  Occurrences   Line Contents
=============================================================
    54     24.8 MiB     24.8 MiB           1       @profile(stream=profile_stream, precision=PROFILE_PRECISION)
    55                                             def findMedianSortedArrays(self, nums1, nums2) -> float:
    56                                                 # Ensure nums1 is the smaller array to minimize binary search domain
    57     24.8 MiB      0.0 MiB           1           if len(nums1) > len(nums2):
    58                                                     nums1, nums2 = nums2, nums1
    59     24.8 MiB      0.0 MiB           1           x, y = len(nums1), len(nums2)
    60                                         
    61     24.8 MiB      0.0 MiB           1           low, high = 0, x
    62     24.8 MiB      0.0 MiB           1           while low <= high:
    63     24.8 MiB      0.0 MiB           1               partitionX = (low + high) // 2
    64     24.8 MiB      0.0 MiB           1               partitionY = (x + y + 1) // 2 - partitionX
    65                                         
    66                                                     # If partitionX is 0 it means nothing is there on left side. Use -inf for maxLeftX
    67     24.8 MiB      0.0 MiB           1               maxLeftX = float('-inf') if partitionX == 0 else nums1[partitionX - 1]
    68                                                     # If partitionX is length of input then there is nothing on right side. Use +inf for minRightX
    69     24.8 MiB      0.0 MiB           1               minRightX = float('inf') if partitionX == x else nums1[partitionX]
    70                                         
    71     24.8 MiB      0.0 MiB           1               maxLeftY = float('-inf') if partitionY == 0 else nums2[partitionY - 1]
    72     24.8 MiB      0.0 MiB           1               minRightY = float('inf') if partitionY == y else nums2[partitionY]
    73                                         
    74     24.8 MiB      0.0 MiB           1               if maxLeftX <= minRightY and maxLeftY <= minRightX:
    75                                                         # The partition is correct
    76     24.8 MiB      0.0 MiB           1                   if (x + y) % 2 == 0:
    77                                                             return (max(maxLeftX, maxLeftY) + min(minRightX, minRightY)) / 2.0
    78                                                         else:
    79                                                             return max(maxLeftX, maxLeftY)
    80                                                     elif maxLeftX > minRightY:  # We are too far on right side for partitionX. Go on left side.
    81                                                         high = partitionX - 1
    82                                                     else:  # We are too far on left side for partitionX. Go on right side.
    83                                                         low = partitionX + 1
\end{minted}
            \end{modelbox}
        \end{minipage}%
    \caption{A case illustration of GPT-4-turbo-preview (FindMedianSortedArrays). Part G: Memory usage profile.}
    \label{fig:ineffective_case_task_code_refine_code_memory}
\end{figure*}

\clearpage
\section*{NeurIPS Paper Checklist}

\begin{enumerate}

\item {\bf Claims}
    \item[] Question: Do the main claims made in the abstract and introduction accurately reflect the paper's contributions and scope?
    \item[] Answer: \answerYes{} 
    \item[] Justification: In this paper, we provide \name to improve the efficiency of LLM-generated code.

\item {\bf Limitations}
    \item[] Question: Does the paper discuss the limitations of the work performed by the authors?
    \item[] Answer: \answerYes{} 
    \item[] Justification: We discuss the limitation of \name in Appendix.

\item {\bf Theory Assumptions and Proofs}
    \item[] Question: For each theoretical result, does the paper provide the full set of assumptions and a complete (and correct) proof?
    \item[] Answer: \answerNA{} 
    \item[] Justification: The paper does not include theoretical results. 

    \item {\bf Experimental Result Reproducibility}
    \item[] Question: Does the paper fully disclose all the information needed to reproduce the main experimental results of the paper to the extent that it affects the main claims and/or conclusions of the paper (regardless of whether the code and data are provided or not)?
    \item[] Answer: \answerYes{} 
    \item[] Justification: We provide all source code of \name in Supplementary file.

\item {\bf Open access to data and code}
    \item[] Question: Does the paper provide open access to the data and code, with sufficient instructions to faithfully reproduce the main experimental results, as described in supplemental material?
    \item[] Answer: \answerYes{} 
    \item[] Justification: The source code has been uploaded into the supplementary file. The dataset used in the paper also has instructions to access.

\item {\bf Experimental Setting/Details}
    \item[] Question: Does the paper specify all the training and test details (e.g., data splits, hyperparameters, how they were chosen, type of optimizer, etc.) necessary to understand the results?
    \item[] Answer: \answerYes{} 
    \item[] Justification: We provide the experimental setting in the main paper and appendix.

\item {\bf Experiment Statistical Significance}
    \item[] Question: Does the paper report error bars suitably and correctly defined or other appropriate information about the statistical significance of the experiments?
    \item[] Answer: \answerYes{} 
    \item[] Justification: Our proposed method is an inference-only approach for LLM and we adopt the greedy-decoding strategy for all of our experiments, making the experiment results of each session consistent.

\item {\bf Experiments Compute Resources}
    \item[] Question: For each experiment, does the paper provide sufficient information on the computer resources (type of compute workers, memory, time of execution) needed to reproduce the experiments?
    \item[] Answer: \answerYes{} 
    \item[] Justification: All information was provided in the Appendix.
    
\item {\bf Code Of Ethics}
    \item[] Question: Does the research conducted in the paper conform, in every respect, with the NeurIPS Code of Ethics \url{https://neurips.cc/public/EthicsGuidelines}?
    \item[] Answer: \answerYes{} 
    \item[] Justification: The research follow the NeurIPS Code of Ethics.

\item {\bf Broader Impacts}
    \item[] Question: Does the paper discuss both potential positive societal impacts and negative societal impacts of the work performed?
    \item[] Answer: \answerYes{} 
    \item[] Justification: We discuss societal impacts in the Appendix.
    
\item {\bf Safeguards}
    \item[] Question: Does the paper describe safeguards that have been put in place for responsible release of data or models that have a high risk for misuse (e.g., pretrained language models, image generators, or scraped datasets)?
    \item[] Answer: \answerNA{} 
    \item[] Justification: the paper poses no such risks.

\item {\bf Licenses for existing assets}
    \item[] Question: Are the creators or original owners of assets (e.g., code, data, models), used in the paper, properly credited and are the license and terms of use explicitly mentioned and properly respected?
    \item[] Answer: \answerYes{} 
    \item[] Justification: We have cite all evaluated models and datasets.

\item {\bf New Assets}
    \item[] Question: Are new assets introduced in the paper well documented and is the documentation provided alongside the assets?
    \item[] Answer: \answerNA{} 
    \item[] Justification: the paper does not release new assets.

\item {\bf Crowdsourcing and Research with Human Subjects}
    \item[] Question: For crowdsourcing experiments and research with human subjects, does the paper include the full text of instructions given to participants and screenshots, if applicable, as well as details about compensation (if any)? 
    \item[] Answer: \answerNA{} 
    \item[] Justification: the paper does not involve crowdsourcing nor research with human subjects.

\item {\bf Institutional Review Board (IRB) Approvals or Equivalent for Research with Human Subjects}
    \item[] Question: Does the paper describe potential risks incurred by study participants, whether such risks were disclosed to the subjects, and whether Institutional Review Board (IRB) approvals (or an equivalent approval/review based on the requirements of your country or institution) were obtained?
    \item[] Answer: \answerNA{} 
    \item[] Justification: the paper does not involve crowdsourcing nor research with human subjects.

\end{enumerate}

\end{document}